\documentclass{aa}

\usepackage{natbib}
\bibpunct{(}{)}{;}{a}{}{,}

\usepackage{graphicx}

\usepackage{txfonts,textcomp}
\usepackage{nameref}

\begin{document} 
\makeatletter
\let\do@linenumber\relax
\let\linenumberpar\relax
\let\linenumbers\relax
\let\runninglinenumbers\relax
\let\resetlinenumber\relax
\let\switchlinenumbers\relax
\makeatother

   \title{Superhydrogenation of indene at low temperatures}

   \author{S. Haid
          \inst{1}
          \and
          K. Gugeler
          \inst{1}
          \and
          J. Kästner\inst{1}
          \and
          D. Campisi
          \inst{1}$^{,}$
          \inst{2}
            }

   \institute{$^1$ Institute for Theoretical Chemistry, University of Stuttgart,
             Pfaffenwaldring 55, 70569 Stuttgart, Germany\\
             $^2$ Current Address: Department of Engineering, University of Perugia, Via Duranti 93, 06125 Perugia, Italy \\
              \email{campisi@theochem.uni-stuttgart.de}
             }

  \abstract 
  {The hydrogenation of polycyclic aromatic hydrocarbons (PAHs) is crucial to understanding molecular hydrogenation formation in the interstellar medium (ISM). This process also helps to elucidate the weakening of the aromatic bonds in PAHs, which may function as a carbon reservoir, facilitating the formation of interstellar complex organic molecules (iCOMs) through top-down chemistry. Tunneling can significantly promote the hydrogenation process in a low-to-moderate temperature range (approximately 10-200 K), which could also be important in warmer regions of the ISM, such as photodissociation regions (PDRs).}
   {We aim to present the hydrogenation sequence of the newly observed PAH molecule, indene, for the first time and clarify the tunneling rule at temperature in PDR and dark molecular-cloud conditions. In addition, we report fit parameters to be utilized in astronomical modeling.}
   {The hydrogenation sequence was studied using simple hydrogenation rules based on tight-binding methods and confirmed by barriers from density functional theory (DFT). The binding energy, activation energies, kinetic rate constants, and tunneling corrections --based on the Bell and Eckart models and supported by the accurate instanton method-- were calculated using DFT. To make our kinetic studies useful to modelers, we implemented a Monte Carlo method-based program to generate and optimize random initial fit parameters ($\alpha$, $\beta$, $\gamma$, and T$_0$) to achieve the statistically best fit.}
   {We find that indene hydrogenation proceeds with saturation of carbon atoms in the pentagonal ring first, followed by hydrogenation of the benzene unit. Indene hydrogenation follows rules similar to those of other PAHs, such as pentacene, coronene, and corannulene, with binding energies for odd-numbered hydrogenation steps ranging from 0.5 to 2 eV and barriers around 0.13 eV for the first, fifth, and seventh hydrogenation steps. The third hydrogenation step is the rate-limiting step, with a barrier of 0.24 eV, similar to what is found for other PAHs. Even-numbered hydrogenation steps have lower barriers and lead to more stable intermediates as a result of radical-radical recombination. The hydrogenation sequence follows a scheme that strongly depends on the PAH's shape, the number of aromatic rings, and the presence of five-membered rings, aiming to preserve the aromaticity as much as possible. Furthermore, we observe that tunneling plays an important role in the hydrogenation of indene at temperatures between 30 and 75 K, which corresponds to the temperatures of dust in PDRs. Finally, our implementation includes fit parameters that reproduce our model with a high degree of accuracy, achieving a static precision of 0.99 (R$^2$) and an RMS error of 10$^{-2}$.}
   {}


   \maketitle

\section{Introduction}
\label{Introduction}
Polycyclic aromatic hydrocarbons (PAHs) are a group of molecules characterized by carbon honeycomb structures bound with hydrogen atoms along their edges. They are commonly observed and detected in the interstellar medium (ISM) and celestial bodies within our solar system \citep{Tielens_2008,Tielens_2013,Burkhardt_2021,Chown_2024}. The formation of PAHs is believed to take place near carbon-rich stars, similarly to processes observed in combustion chemistry \citep{PAHsoot:1989,Cherchneff:1992}. Following their formation, PAHs are expelled into the ISM, where they play an important role in influencing the physics and chemistry of the surrounding environment \citep{Tielens_2013}. Within the ISM, they exist as constituents of carbonaceous grains or as isolated molecules \citep{Tielens_2013,Klarke_2013}. The aromatic nature of PAHs enables them to undergo $\pi$-stacking, leading to the formation of solid macromolecular structures \citep{Rapacioli_2005,Rapacioli_2006}. 

Polycyclic aromatic hydrocarbons are acknowledged for their ability to chemisorb atomic hydrogen onto their carbon lattice, leading to superhydrogenation \citep{Rauls_2008,Jensen_2019,Campisi_2020,Leccese_2023}. This phenomenon is of critical importance within the ISM as it facilitates the formation of molecular hydrogen (H$_2$) through the Eley--Rideal (ER) mechanism \citep{Rauls_2008,Wakelam2017,Foley_2018} or Langmuir--Hinshelwood (LH) mechanism. 
The ER mechanism involves the recombination of atomic hydrogen with one of the chemisorbed hydrogen atoms on PAH surfaces. On the other hand, the LH mechanism operates at surface temperatures that typically range from 10-20 K, where physisorbed atoms migrate across the surface, facilitating encounters, combination, and subsequent extraction as H$_2$ \citep{Wakelam2017}. In regions such as {photodissociation regions} (PDRs), where gas temperatures hover around 300 K and grain temperatures reach approximately 30-75 K at radiation-field equilibrium, it is postulated that the grains function as efficient catalysts for H$_2$ formation. PAHs within PDRs play a pivotal role, notably due to their propensity to form superhydrogenated species, which may contribute to H$_2$ formation through ER mechanisms \citep{Wakelam2017}. 

Theoretical studies have shown that linear neutral PAHs, such as pentacene, exhibit kinetics faster towards the first two hydrogenations compared to coronene \citep{Jensen_2019,Campisi_2022}. However, the third hydrogenation results in a limiting step due to the low binding energy, which can compete with H loss under a radiation field of approximately 6-13 eV \citep{Andrews_2016}. Therefore, doubts arise regarding the efficiency of PAHs in becoming fully hydrogenated in PDRs. On the other hand, H$_2$ formation is also a crucial component in dark and cold molecular clouds with temperatures ranging from 10-50 K, since H$_2$ plays an important role in cloud collapse and the formation of a new protostar \citep{Ferullo_2019,Tielens_2013}. In cold conditions, quantum-tunneling effects must be taken into account to understand the superhydrogenation sequence \citep{Goumans_2010,Goumans_2011,Jelenfi_2023}.

Superhydrogenation not only plays a crucial role in {H$_2$} formation, it also serves to weaken the strong aromatic C--C bonds of PAHs, possibly aiding their fragmentation \citep{Campisi_2020}. This is significant because PAHs lock up a large fraction of cosmic carbon --approximately 20\% \citep{Allamandola_1989,Tielens_2013}-- and thus superhydrogenation may play a key role in weakening strong aromatic bonds and releasing carbon from PAHs. According to the top-down chemistry hypothesis \citep{Tielens_2013}, the action of hydrogen and other radicals could induce PAH fragmentation. Once the daughter molecules of PAHs are formed by fragmentation, they can react with other species to form interstellar complex organic molecules (iCOMs) that might contribute to the formation of prebiotic molecules important for understanding the origin of life on habitable planets \citep{Herbst_2009,Lattelais_2009,Tielens_2013,Campisi_2022}. However, only a few studies have addressed the fragmentation of PAHs due to superhydrogenation \citep{Alliati_2019,Zeyuan_2022}. Theoretical studies have elucidated the formation of ethylene molecules from pyrene cations under UV radiation, involving seven additional hydrogen attachments during superhydrogenation \citep{Zeyuan_2022}. Depending on the position of the hydrogen atoms, only specific configurations of hydrogenation hold the key to favorably releasing ethylene molecules.
 
Recent observations \citep{Burkhardt_2021} conducted by the Green Bank Telescope (GBT) as part of the Hunting for Aromatic Molecules (GOTHAM) survey have identified pure indene (Fig. \ref{Fig:Indene}) as a predominant PAH species within dense, cold, dark molecular clouds of Taurus Molecular Cloud 1 (TMC-1). One might assume that, given the extremely low temperature, PAHs cannot undergo superhydrogenation. However, in principle, light hydrogen atoms could tunnel through a potential barrier, thus allowing the formation of superhydrogenated PAHs at low temperatures \citep{Goumans_2010,Goumans_2011,Indene_2024,Jelenfi_2023}.

In this work, we simulated the superhydrogenation of recently observed indene using density functional theory (DFT), a quantum chemistry method (Section \ref{DFT}). Recent studies have elucidated the first hydrogen addition and extraction process of indene \citep{Indene_2024}, as well as the effect of tunneling at 50 K in the hydrogenation of indene and H$_2$ formation \citep{Jelenfi_2023}, showing that indene could be an efficient catalyst for H$_2$ formation at low temperatures (below 50 K). Here, we compute the binding energies and energy barriers for H attachment up to superhydrogenation, with a total of eight hydrogen atoms attached (Section \ref{Indene's hydrogenation sequence}), and compare the hydrogenation sequence of indene with the sequence of well-known hydrogenated PAHs from the literature (Section \ref{Indene vs. other PAHs' hydrogenation}) such as pentacene, coronene, and corannulene. For species with potential barriers (odd-hydrogenated indene), we compute the rate constants at temperatures ranging from 200 to 10 K using the Bell and Eckart tunneling correction to classical kinetic rate constants (Section \ref{Tunneling and fitting parameters}). Furthermore, we implemented an optimization procedure to accurately derive fit kinetic parameters to be used for astronomical modeling (Section \ref{Implementation}), deriving parameters for the odd-hydrogenated indene. Finally, we draw related astrophysical implications (Section \ref{Astrophysical Implications}) and conclusions (Section \ref{Conclusions}).

\section{Simulations}
\label{Simulations}

\subsection{Density functional theory calculations}
\label{DFT}
We utilized DFT implemented in the ORCA code \citep{Neese_2020} using the M06-2X/pcseg-2 method \citep{Zhao2008,Jensen2014}. This level of theory has already been benchmarked and utilized in previous studies for the hydrogenation of other PAHs, accurately reproducing binding energies and barrier heights compared to coupled cluster calculations \citep{Jensen_2019,Campisi_2022}. For this work, we chose to employ a triple zeta basis set instead of a double zeta one, as reported in previous studies \citep{Campisi_2022}, to minimize the basis set superposition error (BSSE) and to better describe the potential energy surface \citep{TZP_des_2019}, which would otherwise require counterpoise corrections.

Geometrical optimization and frequency calculations to correct the energy at zero point were performed using ORCA within the DL-FIND module \citep{DL-FIND} in the ChemShell environment \citep{met14}. For the optimization of local minima, we employed the Broyden--Fletcher--Goldfarb--Shanno (L-BFGS) algorithm implemented in DL-FIND \citep{Liu1989,DL-FIND}. The tolerance for the geometry optimization was set to 0.00003 a.u. per Bohr radius. The optimization of the transition states was performed using the TS-finding algorithm in ORCA, with tight convergence criteria set to 10$^{-8}$ atomic units. The SCF convergence criteria were set very tight at 10$^{-9}$ atomic units for each calculation.
Rate {constants} were computed via harmonic transition state theory, with all vibrational modes treated as quantum harmonic oscillators. Quantum tunneling was taken into account by one-dimensional tunneling corrections based on the Bell \citep{bel59} and the symmetric Eckart model \citep{eck30} as implemented in DL-FIND \citep{mcc17a}. In each step, barriers for all possible hydrogenation sites were calculated, and the path with the lowest barrier was chosen.

In this study, the term “hydrogenated” refers to an sp$^2$ carbon atom that binds to a hydrogen atom, resulting in an sp$^3$ hybridized carbon atom.

Binding energies were computed as
\begin{eqnarray}
E_\text{b} & = & (E_\text{reag} + E_\text{H}) - E_\text{nH-PAH} \label{eq1}
,\end{eqnarray}
where $E_\text{b}$ is the binding energy; $E_\text{reag}$ is the energy of the reactant, indene or hydrogenated indene; $E_\text{H}$ is the energy of atomic hydrogen; and $E_\text{$n$H-PAH}$ is the energy of the {hydrogenated} product, such as the resulting hydrogenated species. Positive values of $E_\text{b}$ indicate favorable binding energies.

The energy barriers were calculated as 
\begin{eqnarray}
E_\text{act} & = & E_\text{TS} - (E_\text{nH-PAH} + E_\text{H}) \label{eq3}
,\end{eqnarray}
where $E_\text{act}$ is the energy barrier (activation energy) using the energy of H and the reactant at infinite distance ($E_\text{$n$H-PAH}$ and $E_\text{H}$) and $E_\text{TS}$ is the transition-state energy. Each product of the hydrogenation sequence serves as the reactant for the subsequent hydrogenation. Therefore, in Eq. \ref{eq3} $E_\text{$n$H-PAH}$ is only equal to E$_\text{reag}$  for the first hydrogenation. For subsequent hydrogenation steps, the products of the second, fourth, and sixth hydrogenation steps serve as reactants to determine the barriers for the third, fifth, and seventh hydrogenation steps, respectively.

All the energies ($E$) include the zero-point vibrational energy. 
The crossover temperature $T_\text{cross}$ for tunneling, the temperature below which significant tunneling effects are expected, is calculated by $T_\text{cross}=\hbar|\omega_\text{b}|/(2\pi k_\text{B}),$ where $\hbar$ is the reduced Planck constant, $\omega_\text{b}$ is the imaginary frequency associated with the barrier, and $k_\text{B}$ is Boltzmann's constant.

\subsection{Implementation of fit parameters' optimization for astronomical modeling}
\label{Implementation}

In this study, we implemented a procedure to generate and optimize the fit parameters for the rate constants calculated at the DFT level, which can be used for astronomical modeling. We employed the rate {constant} expression provided by \cite{Zheng_2010}. The expression is
\begin{eqnarray}
k & = & \alpha \left( \frac{T}{300 \text{K}} \right)^{\beta} \exp \left( -\frac{\gamma (T + T_0)}{T^2 + T_0^2} \right) \label{eq5}
,\end{eqnarray}
where $k$ is the rate constant in {cm$^3$} s$^{-1}$, $\alpha$ is a fit parameter with the same unit as the rate constant, $\beta$ is the fit parameter describing low temperature behavior, $\gamma$ is the activation-energy fit parameter in kelvin, and $T_0$ is the reference temperature in kelvin.

The present parameters are necessary to reproduce our tunneling rate constants for use in astronomical modeling \citep{Lamberts_2016,Meisner_2017,Lamberts_2017}. Given that the precision of the fit is influenced by the initial guess parameters, which are usually estimated using {physical} intuition, we implemented a procedure to randomly generate a large number of initial guesses using basic Monte Carlo methods \citep{MonteCarlo_1949}, within defined bounds (see Section \ref{sec:Rate_fitting_SI}). Then, these guessed parameters were optimized to provide the best fit.

We used Eckart and Bell corrected rate {constants}, computed at the DFT level, to fit 30 rate {constant} values within a temperature range of 200 K to 10 K. We generated 5000 random initial guesses and optimized them.
The optimization process involved iterative adjustment of the parameters using the sequential least-squares programming (SLSQP) method \citep{kraft1988}. This method optimizes the parameters by minimizing the error of the root mean square (RMS) of the log$_{10}$(k) subject to bounds in the parameters. The convergence criteria were based on the improvement in RMS error and the stability of the parameter values. The tolerance for the fit parameters was set to $10^{-6}$ (units depending on the parameter; {cm$^{3}$} s$^{-1}$ for $\alpha$, unitless for $\beta$, kelvin for $\gamma$ and $T_0$), and the RMS threshold was set to $10^{-4}$. The optimization continued until the changes in parameters were below a set threshold or the improvement in RMS error became negligible. The adequacy of the fit was assessed using the R-squared statistic, which measures the proportion of variance in the observed data that the fit model accounts for. The best parameters were considered those with an R$^2$ value closest to 1.
To test the precision of the optimized parameters obtained, we fed them into Eq. \ref{eq5} to calculate k and reproduce the kinetic curve that we originally computed using DFT (see Section \ref{sec:Rate_fitting_SI}).

\section{Results}
\label{Results}

\subsection{Indene's hydrogenation sequence}
\label{Indene's hydrogenation sequence}
\label{sec:Indene-Hydrogenation}
Indene (Fig. \ref{Fig:Indene}) is a PAH characterized by a six-membered ring and a five-membered ring, where a carbon atom is already hydrogenated to maintain electroneutrality. Fig. \ref{Fig:Indene} shows the {hydrogenation} sequence found{ using simple selection rules and DFT calculations. These rules are based on the tight-binding method reported by \cite{Bonfanti_2011}, which has been validated by \cite{Jensen_2019} and \cite{Campisi_2020} for the case of coronene and pentacene. In this study, the selection rules qualitatively guided us in predicting and identifying the hydrogenation sequence. This was further confirmed by computing and analyzing the trends in binding energies and energy barriers for potential binding sites (see Section \ref{sec:hydrogenation_SI}).} The selection rules applied to indene are further validated by checking the binding energies and energy barriers for each possible hydrogenation site, as reported in Section \ref{sec:hydrogenation_SI}. The selection rules can predict the hydrogenation sequence without requiring calculations. These rules state that a hydrogen atom will react with the carbon atom that has the smallest $\pi$ coordination number (lowest number of sp$^2$ carbon atoms bound to the considered carbon site) and the highest hypercoordination number (second-neighbor sp$^2$ carbons that have the same coordination number as the considered carbon site, without counting hydrogens). Once hydrogen is chemisorbed on the {sp$^2$} carbon site, an unpaired electron is located in the ortho or para position with respect to the hydrogenated carbon atom \citep{Campisi_2020}.

Given that indene already has hydrogenated carbon, the neighboring carbon atom (carbon 1 in Fig. \ref{Fig:Indene}) will be undercoordinated and therefore will have the lowest coordination $\pi$. Hence, carbon 1 is the site with the lowest energy barrier of 0.13 eV (see Section \ref{sec:hydrogenation_SI}), in agreement with the calculations from \cite{Indene_2024, Jelenfi_2023}, and therefore the most favorable carbon site for the first hydrogenation. Once carbon 1 is hydrogenated, the unpaired electron is localized on carbon 2, as shown by the Löwdin spin population analysis \citep{Lowdin:1955}: 0.56 (carbon 2, Fig. \ref{Fig:Indene}), $-0.03$ (carbon 5, Fig. \ref{Fig:Indene}), 0.12 (carbon 6, Fig. \ref{Fig:Indene}). This Löwdin trend has also been confirmed by analysis of the Mulliken spin population \citep{Mulliken_1955}: 0.86 (carbon 2, Fig. \ref{Fig:Indene}), $-0.34$ (carbon 5, Fig. \ref{Fig:Indene}), 0.25 (carbon 6, Fig. \ref{Fig:Indene}).

Once carbons 1 and 2 are hydrogenated, a benzene-like ring remains aromatic and is available to host further hydrogens. Given the presence of a C$_2$ axis and a mirror plane passing through the molecule of carbon 1, carbons 5, 3, and 7 are perfectly equivalent to carbons 6, 4, and 8 (Fig. \ref{Fig:Indene}). Therefore, these carbon atoms have equivalent coordination numbers and might exhibit equivalent reactivity. In a pure benzene ring, all carbon atoms have the same hypercoordination number, since each second neighbor has a coordination of 2 (a carbon site bound to two adjacent carbon atoms). We remind the reader that carbon 2 of indene (Fig. \ref{Fig:Indene}) has been hydrogenated and no longer belongs to the $\pi$ system. Based on our calculations, carbons 3 and 4 (Fig. \ref{Fig:Indene}) have slightly higher binding energy and lower energy barriers of 0.24 eV (see Section \ref{sec:hydrogenation_SI}). Therefore, if we hydrogenate carbon 3 (Fig. \ref{Fig:Indene}) based on the selection rule, the fourth hydrogen goes to the ortho position (carbon 7 or 5, Fig. \ref{Fig:Indene}) or the para position (carbon 4, Fig. \ref{Fig:Indene}). Based on a spin-density analysis performed using Löwdin population analysis, carbon 8 has a population of 0.40, compared to carbons 5 and 7, which have populations of 0.25 and 0.26, respectively (Fig. \ref{Fig:Indene}). The spin population was also confirmed by Mulliken analysis, with carbon 4 having a population of 0.72 and carbons 5 and 7 having populations of 0.44 and 0.45, respectively (Fig. \ref{Fig:Indene}).

Once carbon 4 is hydrogenated, carbons 5 and 7 become equivalent to carbons 6 and 8 due to the presence of a mirror plane that passes perpendicularly through the plane of the molecule from carbon 1 toward the bond formed by carbons 7 and 8 (Fig. \ref{Fig:Indene}), as well as a C$_2$ axis passing through the molecule along the same direction. Hydrogenating carbon 5 requires overcoming an energy barrier of 0.13 eV, while 0.18 eV is needed to hydrogenate carbon 7. At this point, it is clear that the hydrogenation rules no longer apply as we have lost the graphene-like structure. Hence, once carbon 5 is hydrogenated (fifth hydrogenation), carbon 6 has an unpaired electron that cannot be localized and is willing to react with a further hydrogen (sixth hydrogenation) without any barrier (Fig. \ref{Fig:Indene}). Once carbons 5 and 6 are hydrogenated, additional hydrogen (seventh hydrogen) must overcome a barrier of 0.14 eV to chemisorb in carbon 7 or 8 (Fig. \ref{Fig:Indene}), as they are equivalent due to symmetry,  to complete hydrogenation. Once carbon 7 is hydrogenated, the last hydrogen chemisorbs on carbon 8 without any barrier. The binding energies and energy barriers for each hydrogenation site are reported in Section \ref{sec:hydrogenation_SI}.

\subsection{Indene versus other PAHs' hydrogenation.}
\label{Indene vs. other PAHs' hydrogenation}

\begin{figure}
   \centering
   \includegraphics[width=\hsize]{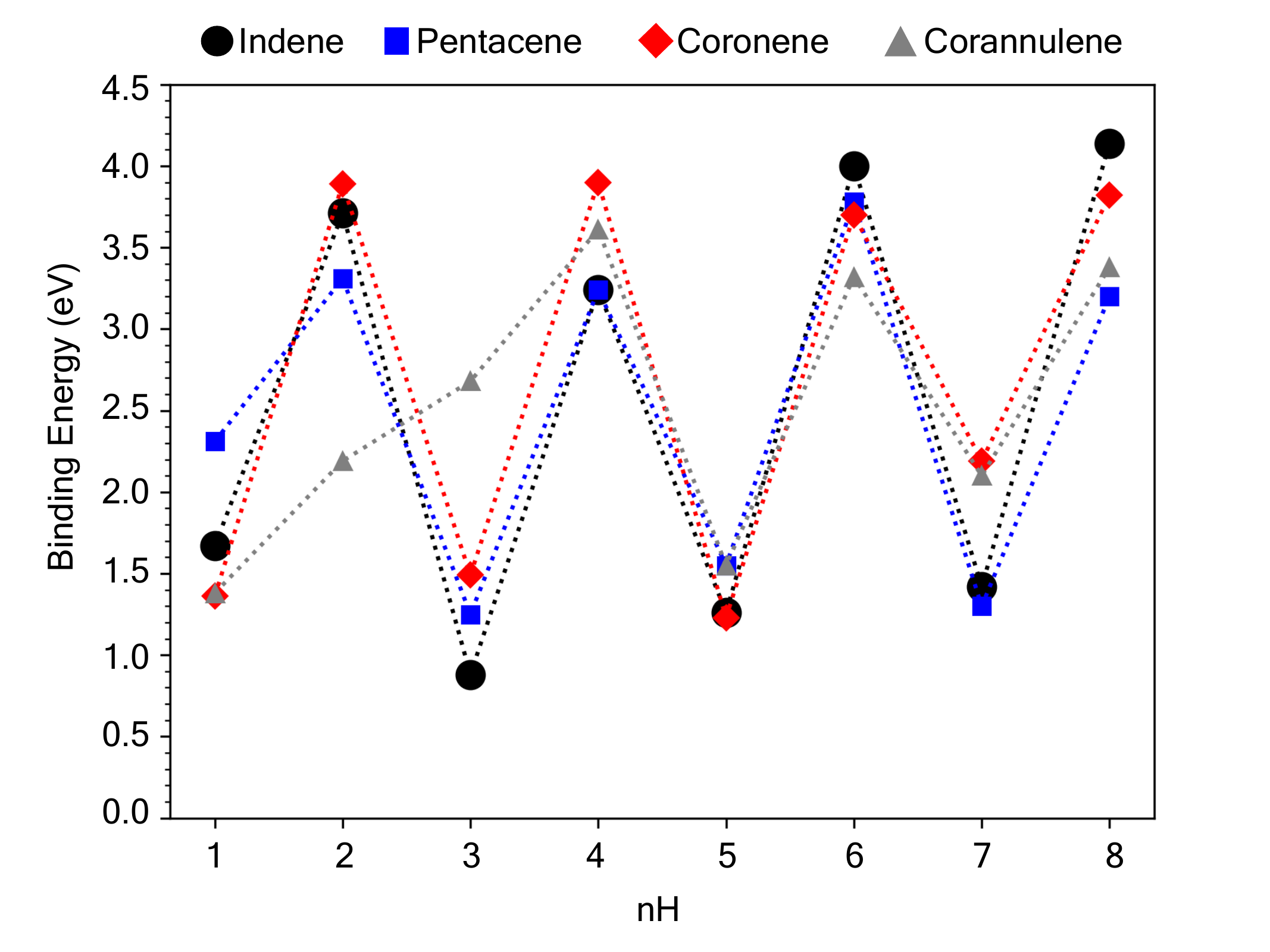}
      \caption{Binding energies for subsequent hydrogenation (nH is the number of added hydrogens) for several PAHs. 
      Black: Indene (this work). Blue: Pentacene \citep{Campisi_2020}. Red:  Coronene (\cite{Jensen_2019}). Gray: Corannulene \citep{Leccese_2023}.
         \label{Fig:BindingEnergies}}
   \end{figure}

\begin{figure}
   \centering
   \includegraphics[width=\hsize]{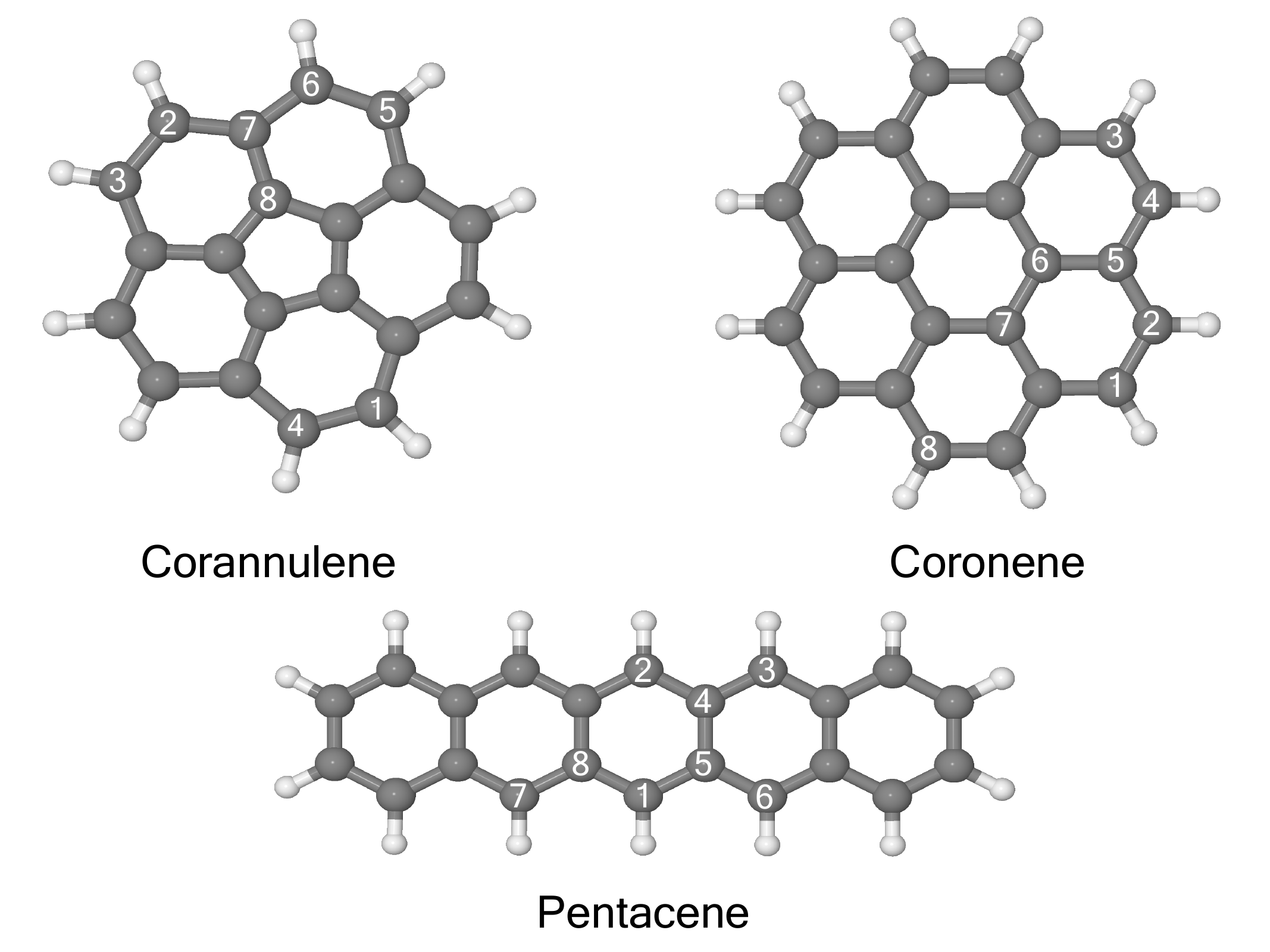}
      \caption{Sequence of first eight hydrogenation steps of corannulene, coronene, and pentacene, as reported by \cite{Leccese_2023}, \cite{Jensen_2019}, and \cite{Campisi_2020}, respectively. Numerical values indicate the hydrogenation step sequence. Gray balls represent carbon atoms, and white balls represent hydrogen atoms.}
         \label{Fig:PAHs_sequence}
   \end{figure}

In Fig. \ref{Fig:BindingEnergies}, we report the trend of the binding energy for indene hydrogenation compared to molecules that have previously been studied, such as pentacene, coronene, and corannulene. In Fig. \ref{Fig:PAHs_sequence}, we report their hydrogenation sequence, limited to eight hydrogenation steps. The differences between these molecules lie in their shapes. Indene, pentacene, and coronene are perfectly planar molecules, while corannulene is a curved PAH because of the presence of a five-membered ring at the center of the lattice.

\begin{figure}
   \centering
   \includegraphics[width=\hsize]{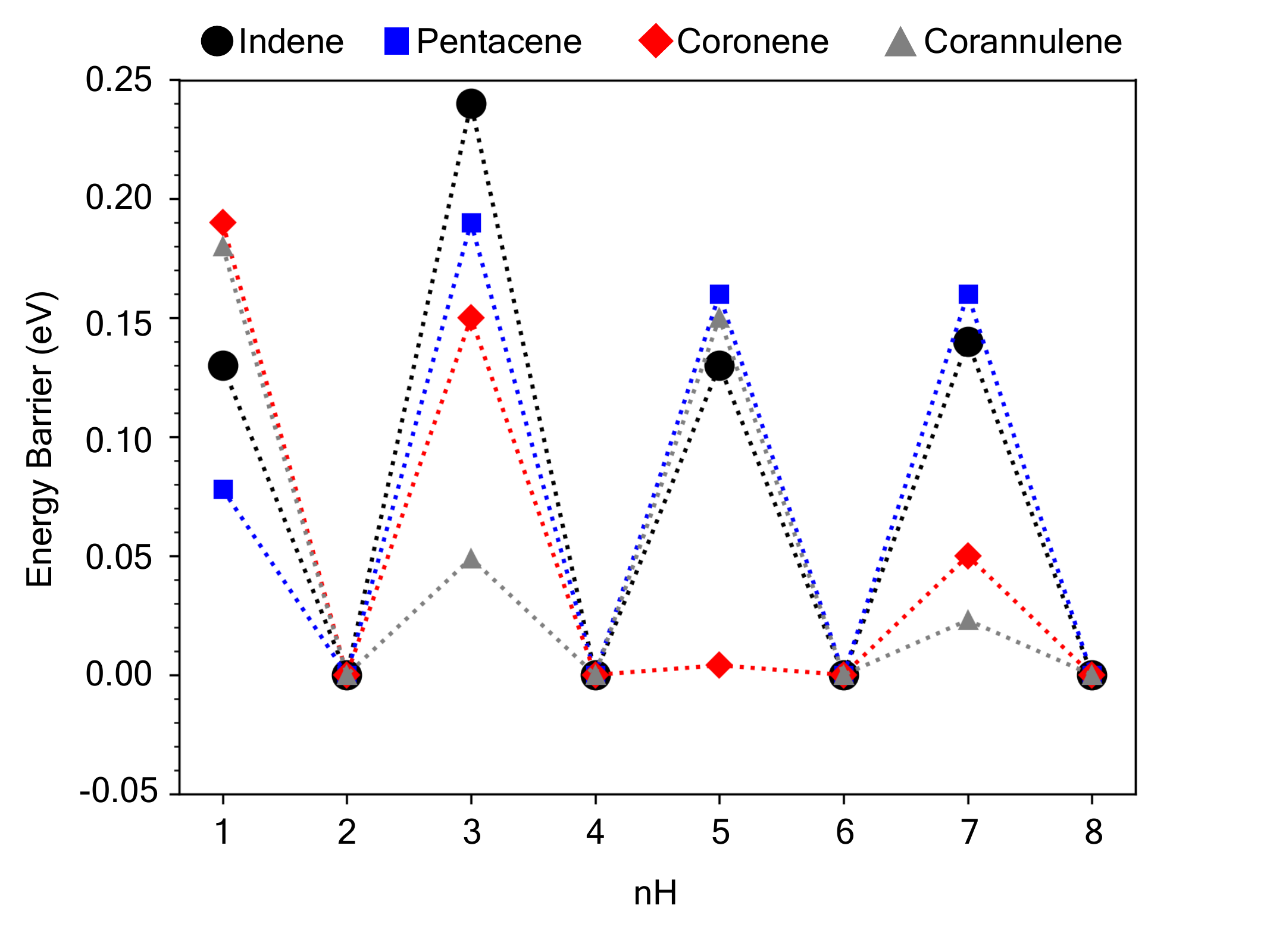}
      \caption{Energy barriers for subsequent hydrogenation (nH is the number of added hydrogens) for several PAHs. 
      Black: Indene (this work). Blue: Pentacene \citep{Campisi_2020}. Red:  Coronene (\cite{Jensen_2019}). Gray: Corannulene \citep{Leccese_2023}.}
         \label{Fig:BarrierEnergies}
   \end{figure}

The binding energy (Fig. \ref{Fig:BindingEnergies}) for the hydrogenation of the mentioned PAHs informs us of their stability \citep{Campisi_2020}. In this context, odd hydrogenations are less stable than even hydrogenations. This is because of the presence of localized unpaired electrons in the PAH lattice. Adding odd hydrogen atoms to carbon sites, they have to overcome barriers (Fig. \ref{Fig:BarrierEnergies}). In contrast, even hydrogenation of carbon atoms results in localized radical atoms that can react with additional hydrogen atoms without such barriers \citep{Rauls_2008,Jensen_2019,Campisi_2020,Leccese_2023}.

\begin{figure}
\centering
\includegraphics[width=\hsize]{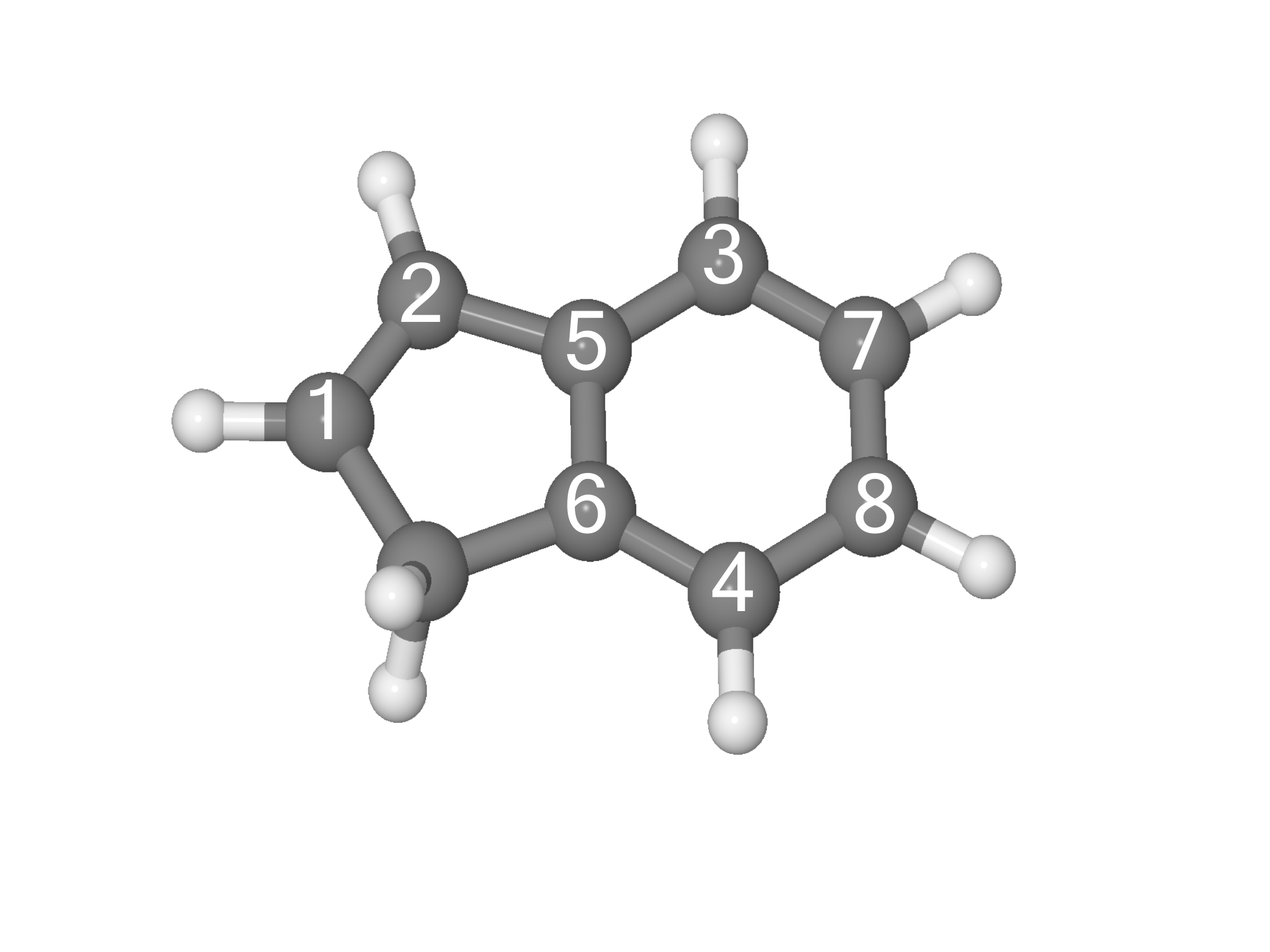}
\caption{Hydrogenation sequence of indene represented by values reported for each carbon atom (gray balls); white atoms are hydrogen atoms.
\label{Fig:Indene}}
\end{figure}

For the first hydrogenation, indene shows greater reactivity toward the initial hydrogen attachment than indene, but lower reactivity than coronene and corannulene, which have the same binding energy. This trend is reflected in their barrier energies, as shown in Fig. \ref{Fig:BarrierEnergies}. As already reported in Section \ref{sec:Indene-Hydrogenation}, for indene, the first hydrogenation occurs on the carbon adjacent to the already-hydrogenated carbon site (carbon 1 in Fig. \ref{Fig:Indene}). For pentacene, this occurs in carbon 1 or 2 of the central ring (Fig. \ref{Fig:PAHs_sequence}), while for corannulene and coronene, this occurs in the duo carbons at their edges (Fig. \ref{Fig:PAHs_sequence}). Since coronene and corannulene have an extended conjuction of the $\pi$-electron system, while linear PAHs such as pentacene and, even less so, indene have higher and less localized electron density, they are more willing to host radical atoms \citep{Cherif_2003}.

For the second hydrogenation, corannulene shows less stability compared to coronene, indene, and pentacene. This is because the center of the corannulene ring is not fully sp$^{2}$ hybridized \citep{Leccese_2023} due to the curvature of the PAH, leading to the localization of unpaired electrons on the edges far from the hydrogenated site. In contrast, for pentacene and coronene, the second hydrogenation occurs at the ortho or para positions. Given that coranulene has a lower aromaticity in comparison to other PAHs, because of the curvature, further hydrogenation compromises the stability because it further reduces the aromaticity. 

For the third hydrogenation, indene shows lower stability and, therefore, higher binding energy compared to all other PAHs. This is because indene has only one aromatic ring, and the third hydrogenation compromises the aromaticity of the system, leading to decreased stability. In contrast, pentacene and coronene lose the aromaticity of two rings during the third hydrogenation. Consequently, they exhibit lower stability compared to corannulene, which has the highest stability for the third hydrogenation. In the case of corannulene, two hydrogens are attached to two different rings, which compromises the aromaticity earlier than those of indene, coronene, and pentacene. Therefore, the third hydrogen attachment in corannulene occurs on a carbon atom whose aromaticity has already been compromised.

For the fourth hydrogenation, coronene (followed by corannulene) shows the greatest stability due to the higher number of conjugated rings remaining intact \citep{Cherif_2003}. In contrast, pentacene exhibits disconnected benzene and naphthalene units, while indene completely loses its aromaticity. 

The fifth hydrogenation exhibits comparable binding energy stability for coronene and indene, as well as for corannulene and pentacene, respectively. However, corannulene and pentacene exhibit higher energy barriers. There is a significant difference in the barrier heights between coronene and indene. Indene and coronene should have the highest barriers given their lower binding energy. In the case of coronene and corannulene, the fifth hydrogenation occurs on two hybridized aromatic sp$^{2}$ carbons, while for pentacene and indene, the fifth hydrogen attaches to an aliphatic sp$^2$ carbon. In pentacene, carbon 5 has a localized double bond with carbon 6, which is conjugated with the benzene unit. Therefore, hydrogenating carbon 5 leads to a loss of stability. Hydrogenating carbon 5 on coronene does not lead to any further loss of aromaticity, while corannulene starts to lose the aromaticity of the third aromatic ring. Therefore, coronene shows lower energy barriers with respect to all PAHs. Indene shows a barrier lower than that of pentacene, and as for pentacene we have a loss of conjugation.

In the sixth hydrogenation, indene appears to be more stable than pentacene, coronene, and corannulene, respectively. In indene, the hydrogenation of the sixth carbon will form a CH=CH unit, while pentacene completes the hydrogenation of two aromatic rings (leaving three aromatic rings). In coronene, the sixth hydrogenation starts to diminish the aromaticity of the central ring, leaving four aromatic rings, while corannulene has only three aromatic rings left.

The seventh hydrogenation reveals that corannulene and indene exhibit stability comparable to that of coronene and lower binding energy. The barrier energies align with the trend of the binding energy (Figs. \ref{Fig:BarrierEnergies} and \ref{Fig:BindingEnergies}). We attribute the higher hydrogenation stability of coronene and corannulene over pentacene and indene to the presence of a greater number of aromatic or conjugated rings.

In the eighth hydrogenation, indene emerges as more stable than coronene, corannulene, and pentacene, respectively. This is because indene undergoes full hydrogenation, losing its aromaticity, while coronene retains three aromatic rings, corannulene retains two six-membered rings and one five-membered ring, and pentacene retains two isolated benzene units.

In general, the trend of activation energies follows the same pattern as the binding energies based on the Bell--Evans--Polanyi principle \citep{Bell_1936,Evans:1938}. Higher stability (higher binding energy) correlates with lower energy barriers. We acknowledge that this principle does not always apply to every hydrogenation case, as the preservation of aromaticity plays a fundamental role in stabilizing the system. A slight divergence is also observed for the seventh hydrogenation of coronene and corannulene, but this difference is within the functional error \citep{Mardirossian_2016}.

\subsection{Tunneling and fit parameters}
\label{Tunneling and fitting parameters}

\begin{table}

\small
    \centering
    \caption{Vibrational adiabatic barrier ($E_\text{act}$), crossover temperatures ($T_\text{cross}$) and absolute barrier frequencies ($|\omega_\text{b}|$ ) for the odd hydrogenations (nH) of indene.}
    \label{tab:barrier_parameters}
    \begin{tabular}{lrrrrr}
        \hline\hline
        nH & $E_\text{act}$ (eV) & $T_\text{cross}$ (K) & $|\omega_\text{b}|$ (cm$^{-1}$)  \\
        \hline
        1H & 0.13 & 147.1 & 642.1\\
        3H & 0.24 & 196.2 & 856.8\\
        5H & 0.13 & 168.8 & 737.3\\
        7H & 0.14 & 160.5 & 700.7\\
        \hline\hline
    \end{tabular}
\end{table}

\begin{figure}
   \centering
   \includegraphics[width=\hsize]{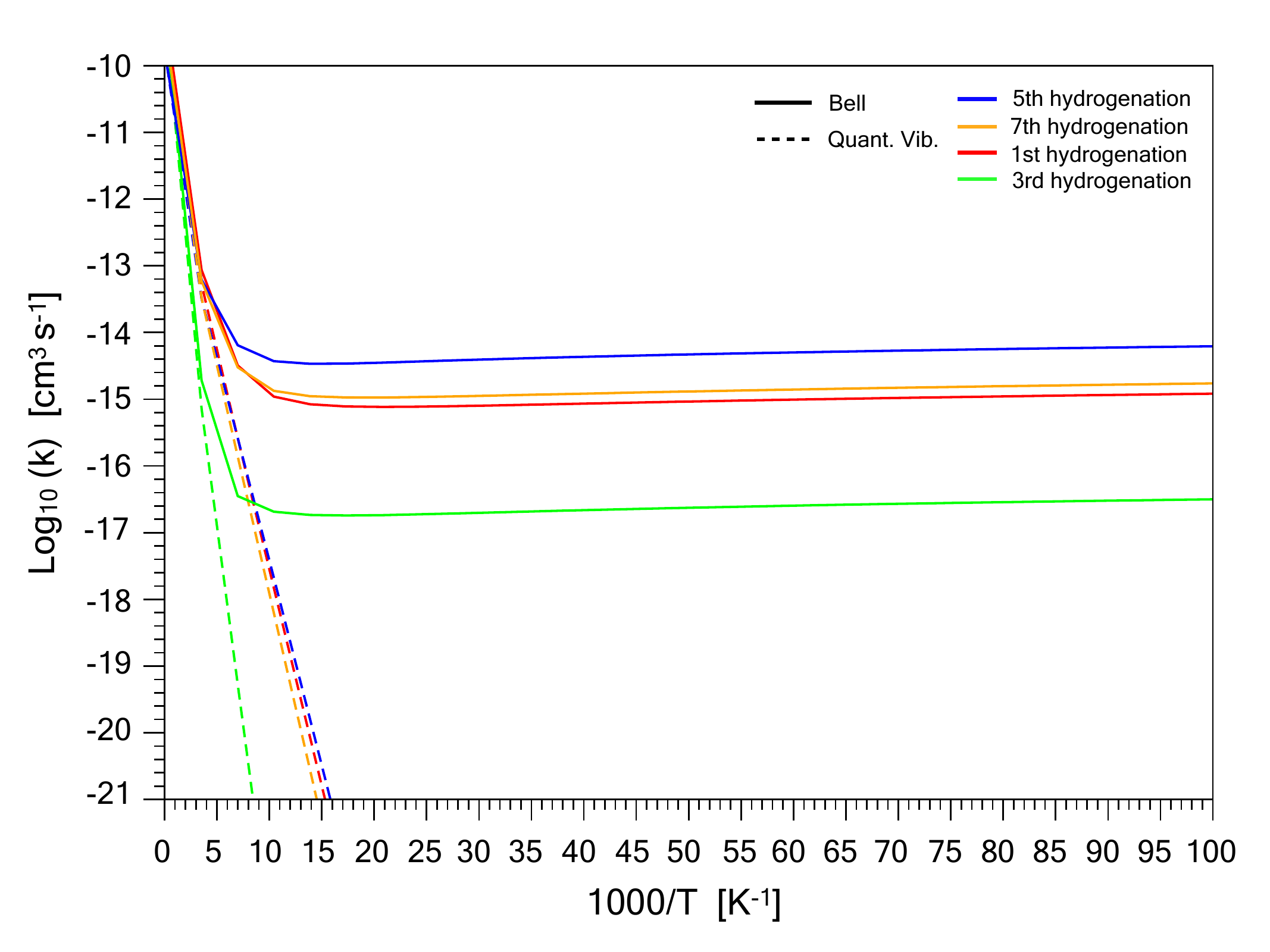}
      \caption{Arrhenius plot shows logarithm {(base 10)} of bimolecular rate constants (Log$_{10}$($k$)), TST rate constants with zero-point energy correction (Quant. Vib.), and Bell tunneling correction (Bell) as a function of 1000 divided by the temperature in kelvin (1000/$T$), for the odd hydrogenation of indene.}
         \label{Fig:Bell_tunneling_bimol}
   \end{figure}

\begin{figure}
   \centering
   \includegraphics[width=\hsize]{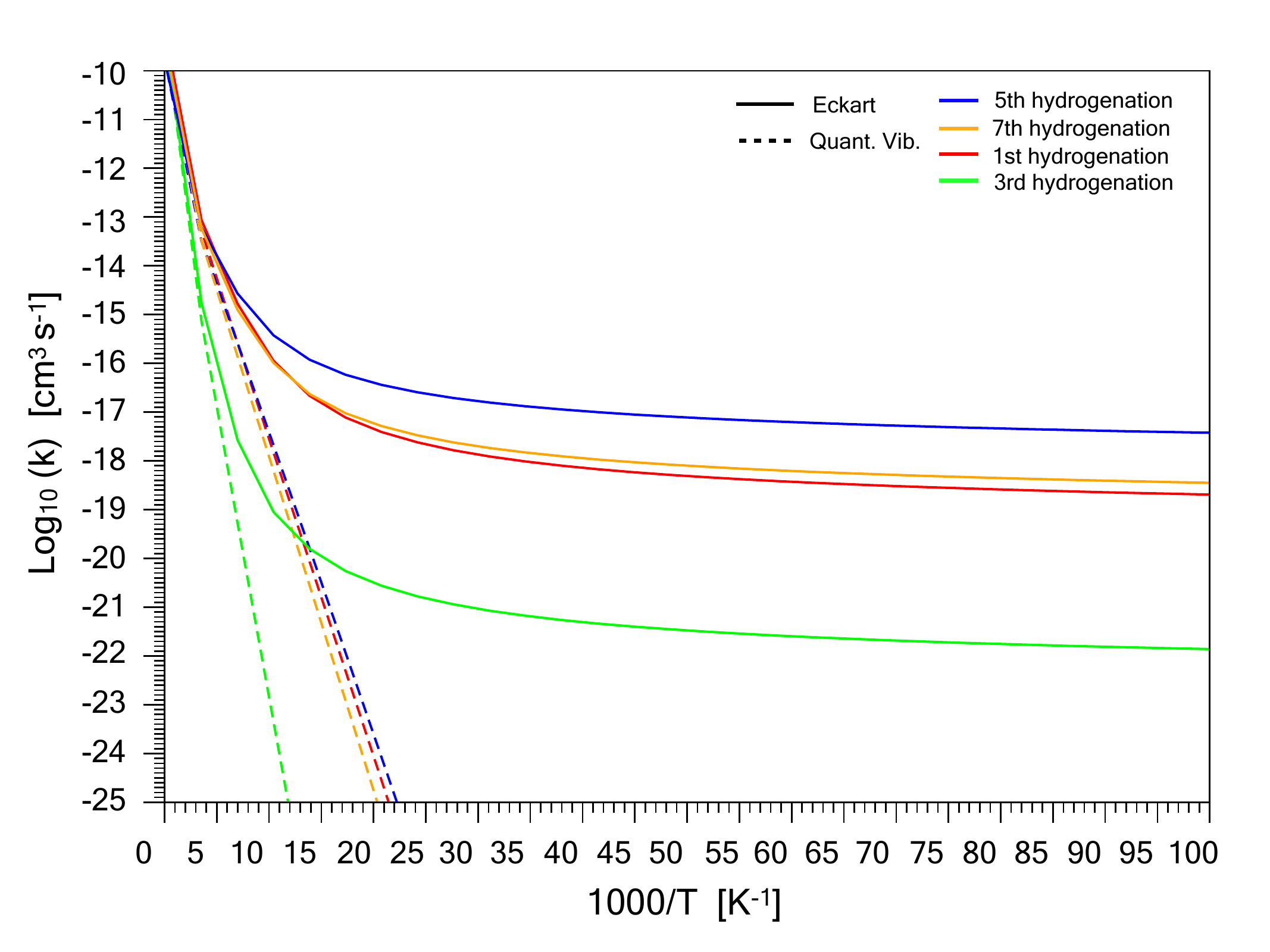}
      \caption{Arrhenius plot shows logarithm (base 10) of bimolecular rate constants (Log$_{10}$($k$)), TST rate constants with zero-point energy correction (Quant. Vib.), and Eckart tunneling correction (Eckart) as a function of 1000 divided by the temperature in kelvin (1000/$T$) for the odd hydrogenation of indene.}
         \label{Fig:Eckart_tunneling_bimol}
   \end{figure}

Figures \ref{Fig:Bell_tunneling_bimol} and \ref{Fig:Eckart_tunneling_bimol} show the Arrhenius plots for the bimolecular reaction of the odd hydrogenations of indene, which present vibrational adiabatic barriers reported in Table \ref{tab:barrier_parameters}.
Here, we focus solely on bimolecular rate constants, which are commonly used to study neutral-neutral reactions in astronomical modeling at temperatures between 10 and 300 K \citep{Wakelam_2012}.

The classical rate {constants} {(Figure \ref{Fig:Bell_tunneling_bimol} and \ref{Fig:Eckart_tunneling_bimol})} do not accurately reproduce the correct hydrogenation rate of indene. This discrepancy arises because the rate constant reaches a plateau at temperatures below 100 K when tunneling corrections are considered. In particular, the crossover temperatures reported in Table \ref{tab:barrier_parameters} for the odd hydrogenation of indene suggest that in dark molecular clouds and in some other regions of the ISM (e.g., PDRs), tunneling could play a major role in the hydrogenation process of PAHs.

Based on Bell tunneling-corrected bimolecular rate constants (Fig. \ref{Fig:Bell_tunneling_bimol}), the third hydrogenation is about two orders of magnitude lower compared to the first and seventh hydrogenation steps, and about three orders of magnitude lower than the fifth hydrogenation. The trend is that the third hydrogenation is lower than the first and the seventh hydrogenations, which present comparable rate constants, and the fifth hydrogenation is the highest.
The slight increase in the rate constant towards low temperature is due to the temperature dependence of the translational partition function in bimolecular reactions.

The Eckart tunneling-corrected rate constants, illustrated in Fig. \ref{Fig:Eckart_tunneling_bimol}, follow a similar trend, but they are lower than the Bell-corrected rate constants. The largest deviation is observed during the third hydrogenation step, where the rate constant at low temperatures is about four orders of magnitude lower than the Bell-corrected rate constants. For the third, fifth, and seventh hydrogenation steps, the difference is around three orders of magnitude compared to the Bell-corrected rate constants. 

From the reported rate constants (Fig. \ref{Fig:Eckart_tunneling_bimol} and \ref{Fig:Bell_tunneling_bimol}), it is clear that the third hydrogenation step, which has a higher energy barrier (Table \ref{tab:barrier_parameters}), is the limiting step of the reaction. At low temperatures (below 200 K), the Eckart rate constants are generally close to the most accurate instanton rate constants (Tables \ref{tab:Instanton_fit_1H} - \ref{tab:Instanton_fit_7H}) and are typically lower by about one order of magnitude \citep{mcc17a}. On the other hand, Bell rate constants are shown to overestimate those compared to Eckart (Tables \ref{tab:fitted_eckart_1H}, \ref{tab:fitted_eckart_3H}, \ref{tab:fitted_eckart_5H}, \ref{tab:fitted_eckart_7H} and \ref{tab:AsymEckart_fit_1H} - \ref{tab:AsymEckart_fit_7H}) and instanton ones (Tables \ref{tab:Instanton_fit_1H} - \ref{tab:Instanton_fit_7H}). This has also been observed by \cite{mcc17a}, which showed that Bell overestimates the rate constant relative to the instanton one by about two orders of magnitude for the reaction of OH with H$_2$ at a temperature of 150 K.

To better understand the tunneling rate {constants}, we fit the tunneling curves (Fig. \ref{Fig:Bell_tunneling_bimol} and \ref{Fig:Eckart_tunneling_bimol}), obtaining the fit parameters based on Eq. \ref{eq5}  reported in Section \ref{Implementation}. These fit parameters not only allow us to make our results available to modelers, they also help us understand the tunneling effect in the hydrogenation of indene. Table \ref{tab:fitting_parameters} shows the fit parameters computed using our implementation.

The parameters $\alpha$, which are related to the behavior of the rate constant, show higher rates for the Bell model compared to the Eckart model, in agreement with the trends reported in Figs. \ref{Fig:Bell_tunneling_bimol} and \ref{Fig:Eckart_tunneling_bimol}. For the Bell model, the parameter $\alpha$ does not vary significantly between hydrogenation steps (with differences of less than an order of magnitude), while in the Eckart model, the parameter $\alpha$ for the third hydrogenation step is an order of magnitude lower than for the first, fifth, and seventh hydrogenations.
The $\beta$ parameters, associated with the low temperature behavior of the rate constants, yield higher values for the Eckart model and lower values for the Bell model (about half the size). In both models, the the trend computed for the hydrogenation steps shows higher $\beta$ values for the third hydrogenation, followed by the first, fifth, and seventh hydrogenations.
The $\gamma$ parameters, fit for the barrier height, indicate a larger barrier for the Bell model compared to the Eckart one, suggesting that the Eckart model exhibits more pronounced tunneling effects. 
The $\gamma$ values for both models show higher values for the third hydrogenation (indicating a higher energy barrier for this step), while the differences for the first and seventh hydrogenations are minor due to similar barrier heights (Table \ref{tab:barrier_parameters}).
The reference temperature-fit parameter, $T_0$, similarly to the $\gamma$ parameters, results in larger values for the Bell model compared to the Eckart one, indicating that the reduced reference temperature in the Eckart model better captures the tunneling effects.

\begin{table}
    \centering
    \caption{{Fit parameters.}}
    \label{tab:fitting_parameters}
    \begin{tabular}{@{}cccccc@{}}
        \hline\hline
         nH & \textbf{Model} & $\boldsymbol{\alpha}$ & $\boldsymbol{\beta}$ & $\boldsymbol{\gamma}$ & $\boldsymbol{T_0}$ \\
        \hline
        1H & Bell & $9.86$ x $10^{-11}$ & $0.54$ & $1652$ & $181$ \\
        & Eckart & $6.44$ x $10^{-13}$ & $2.00$ & $525$ & $72$ \\
        \hline
        3H & Bell & $4.73$ x $10^{-11}$ & $0.65$ & $2549$ & $218$ \\
        & Eckart & $7.75$ x $10^{-14}$ & $2.43$ & $975$ & $90$ \\
        \hline
        5H & Bell & $4.78$ x $10^{-11}$ & $0.28$ & $1661$ & $215$ \\
        & Eckart & $4.99$ x $10^{-13}$ & $1.72$ & $487$ & $89$ \\
        \hline
        7H & Bell & $6.74$ x $10^{-11}$ & $0.47$ & $1733$ & $199$ \\
        & Eckart & $5.23$ x $10^{-13}$ & $1.97$ & $538$ & $79$ \\
        \hline\hline
    \end{tabular}
     \tablefoot{$\alpha$, in cm$^3$ s$^{-1}$, $\beta$ (unit-less), $\gamma$ and $T_0$, in kelvin, were obtained using our implementation for the odd hydrogenation (nH) of indene with bimolecular Eckart and Bell tunneling-rate constants.}
\end{table}

\section{Astrophysical implications}
\label{Astrophysical Implications}
In PDRs, newly formed stars irradiate their surroundings with a radiation field ranging from about 6 to 16.3 eV \citep{Hollenbach_1999,Wakelam2017}. This radiation is sufficient to warm the gas in nearby molecular clouds but not {intense} enough to ionize atomic hydrogen \citep{Hollenbach_1999,Wakelam2017}. H$_2$ formation is believed to be efficiently catalyzed by dust grains \citep{Habart_2004}. 

The chemisorption of hydrogen atoms in PAHs holds promise because it could catalyze H$_2$ formation through the extraction of a second hydrogen atom (ER mechanism), which chemisorbs over the first, thus extracting H$_2$ \citep{Rauls_2008,Wakelam2017}. The radiation field mentioned can photodissociate molecular species in the gas phase, raising the temperature to about 200-500 K due to irradiation \citep{Montillaud_2013,Andrews_2016}. However, in dust grains, the dust reaches thermal equilibrium with the radiation field, resulting in temperatures ranging from 30 K to 75 K \citep{Hollenbach_1999}. Indene can undergo superhydrogenation at temperatures between 30 K and 75 K if adsorbed on a grain surface, which is below the crossover temperature we computed for odd hydrogenation.

We particularly observed tunneling effects below approximately 100 K.  
The computed Eckart rate constants agree well with those obtained for the hydrogenation of pyrene using instanton theory \citep{Goumans_2011}. In particular, the first, fifth, and seventh hydrogenation steps fall within the same range as those predicted by \cite{Goumans_2011} (~10$^{-16}$ - 10$^{-18}$ cm$^3$ s$^{-1}$), while the third hydrogenation step is the rate-limiting step of the reaction, with values around 10$^{-21}$ cm$^3$ s$^{-1}$. Therefore, we anticipate efficient first-step hydrogenation of indene at a dust temperature of 50 K in PDRs, driven by the combined effects of relatively high temperatures and quantum tunneling. In dark molecular clouds (around 10 K), tunneling is expected to dominate over thermal processes. However, similarly to coronene \citep{Andrews_2016} and pentacene \citep{Campisi_2020}, we do not expect indene hydrogenation to proceed beyond the second hydrogenation step, even when tunneling effects are taken into account.

\section{Conclusions}
\label{Conclusions}
In this work, we focused solely on investigating the hydrogenation sequence of indene PAH, which was recently observed in its non-hydrogenated form \citep{Burkhardt_2021}, at temperatures below the crossover temperatures (Table \ref{tab:barrier_parameters}). Future studies will need to clarify the process of molecular hydrogenation extraction, which is beyond the scope of this paper.

We find that indene, similarly to other studied PAHs such as pentacene and coronene \citep{Jensen_2019,Campisi_2020}, follows the hydrogenation rules observed in graphene-like structures \citep{Bonfanti_2011}. We identified indene hydrogenation sequences, as shown in Fig. \ref{Fig:Indene}, where the first two hydrogen atoms chemisorb on the pentagon ring, while the remaining benzene ring is the last part of the indene to be hydrogenated.

As in previous studies on PAHs such as pentacene \citep{Campisi_2020} and coronene \citep{Jensen_2019}, we found that the third hydrogenation is the limiting step due to the presence of a barrier greater than 0.20 eV. The first, fifth, and seventh hydrogenation steps present comparable barriers of about 0.15 eV. Odd hydrogenation events of indene present binding energies between 0.5 and 2 eV, whereas even hydrogenation events present binding energies between 3 and 4.5 eV. We compared indene hydrogenation with the first eight hydrogenations of pentacene, coronene, and corannulene.

We find that the hydrogenation sequence strongly depends on the shape of the PAH. The divergences found among these PAHs depend on their planarity, the presence of five-membered rings, and the conformation of the molecule; for example, a linear sequence of hexagonal rings such as pentacene or a condensed sequence of hexagonal rings such as coronene. The differences in the binding energies and activation energies of various PAHs strongly depend on the number of aromatic rings and whether hydrogenation occurs on an aromatic carbon or on an aliphatic one. Therefore, the hydrogenation sequence for all PAHs is driven by the preservation of the aromatic rings. 

In addition, for the superhydrogenation of indene, we computed the bimolecular tunneling rate constants based on the Bell and Eckart models and find fit parameters to be utilized for astronomical modeling. Based on kinetic rate constants, tunneling dominates at temperatures below about 100 K. The third hydrogenation, as expected from the barrier height, is the slowest process, while the {fifth} hydrogenation is the fastest hydrogenation process. The third hydrogenation occurs on an aromatic ring, whereas the first and fifth hydrogenations occur on an aliphatic one. We find that the Eckart fit parameters show reduced barriers and that all parameters are significantly affected by tunneling effects compared to the Bell fit parameters for the odd hydrogenations of indene.

Our calculations show that, considering an indene PAH on a grain under PDR conditions, tunneling plays an important role in explaining the hydrogenation sequence at dust temperatures of 30-75 K.  Therefore, tunneling under PDR conditions needs to be taken into account; it could play an important role in the hydrogenation process under dust-grain conditions. However, the third hydrogenation remains the limiting hydrogenation step regardless of the shape of the PAH, with the exception of corrannulene, which exhibits a curved structure \citep{Leccese_2023} even when tunneling effects are considered. The present results provide promising fit parameters that can be used in future modeling to clarify the efficiency of PAH hydrogenation.

The hydrogenation of PAHs holds promise not only because it could act as a catalyst for H$_2$ in the ISM, but also because it could elucidate a starting pathway to weaken their strong aromatic bonds. This could lead to the release of carbon atoms, which could act as precursors of iCOMs, potentially explaining the carbon origin in the so-called Solar System organic inventory found in meteorites on Earth \citep{Sephton_Botta_2005}. 

\section*{Data availability}
{The input and output files from the DFT calculations, the corresponding molecular geometries, and the Monte Carlo–based optimization and fitting code, along with the associated fitted parameters and validation tests, are all available in digital format in the Zenodo repository \citep{Campisi_2024}.} 

\begin{acknowledgements}
D.C. conceived and supervised the project and developed and performed the Monte Carlo parameter fitting and optimization code. Both S.H. (supervised by K.G. and D.C.) and D.C. carried out the DFT calculations. J.K. contributed to some of the analysis. D.C. wrote the manuscript; all authors revised it and approved the final version.
D.C. acknowledges the Alexander von Humboldt Foundation for funding.
The authors acknowledge the state of Baden-Württemberg through bwHPC and the German Research Foundation (DFG) through grant no. INST 40/575-1 FUGG (JUSTUS 2 cluster).
\end{acknowledgements}

\bibliographystyle{aa} 
\bibliography{aa}

\begin{thebibliography}{60}
\expandafter\ifx\csname natexlab\endcsname\relax\def\natexlab#1{#1}\fi

\bibitem[{{Allamandola} {et~al.}(1989){Allamandola}, {Tielens}, \& {Barker}}]{Allamandola_1989}
{Allamandola}, L.~J., {Tielens}, A.~G.~G.~M., \& {Barker}, J.~R. 1989, \apjs, 71, 733

\bibitem[{Alliati {et~al.}(2019)Alliati, Donaghy, Tu, \& Bradley}]{Alliati_2019}
Alliati, M., Donaghy, D., Tu, X., \& Bradley, J.~W. 2019, J. Phys. Chem. A, 123, 2107

\bibitem[{{Andrews, H.} {et~al.}(2016){Andrews, H.}, {Candian, A.}, \& {Tielens, A. G. G. M.}}]{Andrews_2016}
{Andrews, H.}, {Candian, A.}, \& {Tielens, A. G. G. M.} 2016, A\&A, 595, A23

\bibitem[{Bell(1959)}]{bel59}
Bell, R.~P. 1959, Trans. Faraday Soc., 55, 1

\bibitem[{Bell \& Hinshelwood(1936)}]{Bell_1936}
Bell, R.~P. \& Hinshelwood, C.~N. 1936, Proc. R. Soc. Lond. A Math. Phys. Sci., 154, 414

\bibitem[{Bonfanti {et~al.}(2011)Bonfanti, Casolo, Tantardini, Ponti, \& Martinazzo}]{Bonfanti_2011}
Bonfanti, M., Casolo, S., Tantardini, G.~F., Ponti, A., \& Martinazzo, R. 2011, J. Chem. Phys., 135, 164701

\bibitem[{Burkhardt {et~al.}(2021)Burkhardt, Lee, Changala, Shingledecker, Cooke, Loomis, Wei, Charnley, Herbst, McCarthy, \& McGuire}]{Burkhardt_2021}
Burkhardt, A.~M., Lee, K. L.~K., Changala, P.~B., {et~al.} 2021, ApJ Letters, 913, L18

\bibitem[{Campisi {et~al.}(2022)Campisi, Lamberts, Dzade, Martinazzo, ten Kate, \& Tielens}]{Campisi_2022}
Campisi, D., Lamberts, T., Dzade, N.~Y., {et~al.} 2022, ACS Earth and Space Chem., 6, 2009

\bibitem[{Campisi {et~al.}(2020)Campisi, Simonsen, Thrower, Jaganathan, Hornekær, Martinazzo, \& Tielens}]{Campisi_2020}
Campisi, D., Simonsen, F. D.~S., Thrower, J.~D., {et~al.} 2020, Phys. Chem. Chem. Phys., 22, 1557

\bibitem[{{Cherchneff} {et~al.}(1992){Cherchneff}, {Barker}, \& {Tielens}}]{Cherchneff:1992}
{Cherchneff}, I., {Barker}, J.~R., \& {Tielens}, A. G.~G.~M. 1992, \apj, 401, 269

\bibitem[{{Chown, R.} {et~al.}(2024){Chown, R.}, {Sidhu, A.}, {Peeters, E.}, {Tielens, Alexander G. G. M.}, {Cami, Jan}, {Berné, Olivier}, {Habart, Emilie}, {Alarcón, Felipe}, {Canin, Amélie}, {Schroetter, Ilane}, {Trahin, Boris}, {Van De Putte, Dries}, {Abergel, Alain}, {Bergin, Edwin A.}, {Bernard-Salas, Jeronimo}, {Boersma, Christiaan}, {Bron, Emeric}, {Cuadrado, Sara}, {Dartois, Emmanuel}, {Dicken, Daniel}, {El-Yajouri, Meriem}, {Fuente, Asunción}, {Goicoechea, Javier R.}, {Gordon, Karl D.}, {Issa, Lina}, {Joblin, Christine}, {Kannavou, Olga}, {Khan, Baria}, {Lacinbala, Ozan}, {Languignon, David}, {Le Gal, Romane}, {Maragkoudakis, Alexandros}, {Meshaka, Raphael}, {Okada, Yoko}, {Onaka, Takashi}, {Pasquini, Sofia}, {Pound, Marc W.}, {Robberto, Massimo}, {Röllig, Markus}, {Schefter, Bethany}, {Schirmer, Thiébaut}, {Vicente, Sílvia}, {Wolfire, Mark G.}, {Zannese, Marion}, {Aleman, Isabel}, {Allamandola, Louis}, {Auchettl, Rebecca}, {Baratta, Giuseppe Antonio}, {Bejaoui, Salma}, {Bera, Partha P.},
  {Black, John H.}, {Boulanger, François}, {Bouwman, Jordy}, {Brandl, Bernhard}, {Brechignac, Philippe}, {Brünken, Sandra}, {Buragohain, Mridusmita}, {Burkhardt, Andrew}, {Candian, Alessandra}, {Cazaux, Stéphanie}, {Cernicharo, Jose}, {Chabot, Marin}, {Chakraborty, Shubhadip}, {Champion, Jason}, {Colgan, Sean W. J.}, {Cooke, Ilsa R.}, {Coutens, Audrey}, {Cox, Nick L. J.}, {Demyk, Karine}, {Meyer, Jennifer Donovan}, {Foschino, Sacha}, {García-Lario, Pedro}, {Gavilan, Lisseth}, {Gerin, Maryvonne}, {Gottlieb, Carl A.}, {Guillard, Pierre}, {Gusdorf, Antoine}, {Hartigan, Patrick}, {He, Jinhua}, {Herbst, Eric}, {Hornekaer, Liv}, {Jäger, Cornelia}, {Janot-Pacheco, Eduardo}, {Kaufman, Michael}, {Kemper, Francisca}, {Kendrew, Sarah}, {Kirsanova, Maria S.}, {Klaassen, Pamela}, {Kwok, Sun}, {Labiano, Álvaro}, {Lai, Thomas S.-Y.}, {Lee, Timothy J.}, {Lefloch, Bertrand}, {Le Petit, Franck}, {Li, Aigen}, {Linz, Hendrik}, {Mackie, Cameron J.}, {Madden, Suzanne C.}, {Mascetti, Joëlle}, {McGuire, Brett A.}, {Merino,
  Pablo}, {Micelotta, Elisabetta R.}, {Misselt, Karl}, {Morse, Jon A.}, {Mulas, Giacomo}, {Neelamkodan, Naslim}, {Ohsawa, Ryou}, {Omont, Alain}, {Paladini, Roberta}, {Palumbo, Maria Elisabetta}, {Pathak, Amit}, {Pendleton, Yvonne J.}, {Petrignani, Annemieke}, {Pino, Thomas}, {Puga, Elena}, {Rangwala, Naseem}, {Rapacioli, Mathias}, {Ricca, Alessandra}, {Roman-Duval, Julia}, {Roser, Joseph}, {Roueff, Evelyne}, {Rouillé, Gaël}, {Salama, Farid}, {Sales, Dinalva A.}, {Sandstrom, Karin}, {Sarre, Peter}, {Sciamma-O’Brien, Ella}, {Sellgren, Kris}, {Shenoy, Sachindev S.}, {Teyssier, David}, {Thomas, Richard D.}, {Togi, Aditya}, {Verstraete, Laurent}, {Witt, Adolf N.}, {Wootten, Alwyn}, {Zettergren, Henning}, {Zhang, Yong}, {Zhang, Ziwei E.}, \& {Zhen, Junfeng}}]{Chown_2024}
{Chown, R.}, {Sidhu, A.}, {Peeters, E.}, {et~al.} 2024, A\&A, 685, A75

\bibitem[{Eckart(1930)}]{eck30}
Eckart, C. 1930, Phys. Rev., 35, 1303

\bibitem[{Evans \& Polanyi(1938)}]{Evans:1938}
Evans, M.~G. \& Polanyi, M. 1938, Trans. Faraday Soc., 34, 11

\bibitem[{Ferullo {et~al.}(2019)Ferullo, Zubieta, \& Belelli}]{Ferullo_2019}
Ferullo, R.~M., Zubieta, C.~E., \& Belelli, P.~G. 2019, Phys. Chem. Chem. Phys., 21, 12012

\bibitem[{Foley {et~al.}(2018)Foley, Cazaux, Egorov, Boschman, Hoekstra, \& Schlathölter}]{Foley_2018}
Foley, N., Cazaux, S., Egorov, D., {et~al.} 2018, MNRAS, 479, 649

\bibitem[{{Frenklach} \& {Feigelson}(1989)}]{PAHsoot:1989}
{Frenklach}, M. \& {Feigelson}, E.~D. 1989, \apj, 341, 372

\bibitem[{Goumans(2011)}]{Goumans_2011}
Goumans, T. P.~M. 2011, MNRAS, 415, 3129

\bibitem[{Goumans \& Kästner(2010)}]{Goumans_2010}
Goumans, T. P.~M. \& Kästner, J. 2010, Angew. Chem. Int. Ed., 49, 7350

\bibitem[{{Habart, E.} {et~al.}(2004){Habart, E.}, {Boulanger, F.}, {Verstraete, L.}, {Walmsley, C. M.}, \& {Pineau des Forêts, G.}}]{Habart_2004}
{Habart, E.}, {Boulanger, F.}, {Verstraete, L.}, {Walmsley, C. M.}, \& {Pineau des Forêts, G.} 2004, A\&A, 414, 531

\bibitem[{Haid {et~al.}(2025)Haid, Gugeler, Kästner, \& Campisi}]{Campisi_2024}
Haid, S., Gugeler, G., Kästner, J., \& Campisi, D. 2025, Zenodo, https://doi.org/10.5281/zenodo.15459177

\bibitem[{Herbst \& van Dishoeck(2009)}]{Herbst_2009}
Herbst, E. \& van Dishoeck, E.~F. 2009, Annu. Rev. Astron. Astrophys., 47, 427

\bibitem[{Hollenbach \& Tielens(1999)}]{Hollenbach_1999}
Hollenbach, D.~J. \& Tielens, A. G. G.~M. 1999, Rev. Mod. Phys., 71, 173

\bibitem[{Jelenfi {et~al.}(2023)Jelenfi, Schneiker, Tajti, G., \& G.}]{Jelenfi_2023}
Jelenfi, D.~P., Schneiker, A., Tajti, A., G., M., \& G., T. 2023, Mol. Phys., 121, e2142168

\bibitem[{Jensen(2014)}]{Jensen2014}
Jensen, F. 2014, J. Chem. Theory Comput., 10, 1074

\bibitem[{Jensen {et~al.}(2019)Jensen, Leccese, Simonsen, Skov, Bonfanti, Thrower, Martinazzo, \& Hornekær}]{Jensen_2019}
Jensen, P.~A., Leccese, M., Simonsen, F. D.~S., {et~al.} 2019, MNRAS, 486, 5492

\bibitem[{Johnston \& Heicklen(1962)}]{Asym_Eckart}
Johnston, H.~S. \& Heicklen, J. 1962, J. Phys. Chem., 66, 532

\bibitem[{{Klærke, B.} {et~al.}(2013){Klærke, B.}, {Toker, Y.}, {Rahbek, D. B.}, {Hornekær, L.}, \& {Andersen, L. H.}}]{Klarke_2013}
{Klærke, B.}, {Toker, Y.}, {Rahbek, D. B.}, {Hornekær, L.}, \& {Andersen, L. H.} 2013, A\&A, 549, A84

\bibitem[{Kraft(1988)}]{kraft1988}
Kraft, D. 1988, A Software Package for Sequential Quadratic Programming, Deutsche Forschungs- und Versuchsanstalt f{\"u}r Luft- und Raumfahrt K{\"o}ln: Forschungsbericht (Wiss. Berichtswesen d. DFVLR)

\bibitem[{Kästner {et~al.}(2009)Kästner, Carr, Keal, Thiel, Wander, \& Sherwood}]{DL-FIND}
Kästner, J., Carr, J.~M., Keal, T.~W., {et~al.} 2009, J. Phys. Chem. A, 113, 11856

\bibitem[{Lamberts {et~al.}(2016)Lamberts, Samanta, Köhn, \& Kästner}]{Lamberts_2016}
Lamberts, T., Samanta, P.~K., Köhn, A., \& Kästner, J. 2016, Phys. Chem. Chem. Phys., 18, 33021

\bibitem[{{Lamberts, T.} {et~al.}(2017){Lamberts, T.}, {Fedoseev, G.}, {Kästner, J.}, {Ioppolo, S.}, \& {Linnartz, H.}}]{Lamberts_2017}
{Lamberts, T.}, {Fedoseev, G.}, {Kästner, J.}, {Ioppolo, S.}, \& {Linnartz, H.} 2017, A\&A, 599, A132

\bibitem[{Lattelais {et~al.}(2009)Lattelais, Pauzat, Ellinger, \& Ceccarelli}]{Lattelais_2009}
Lattelais, M., Pauzat, F., Ellinger, Y., \& Ceccarelli, C. 2009, ApJ, 696, L133

\bibitem[{Leccese {et~al.}(2023)Leccese, Jaganathan, Slumstrup, Thrower, Hornekær, \& Martinazzo}]{Leccese_2023}
Leccese, M., Jaganathan, R., Slumstrup, L., {et~al.} 2023, MNRAS, 519, 5567

\bibitem[{Liu \& Nocedal(1989)}]{Liu1989}
Liu, D.~C. \& Nocedal, J. 1989, Math. Program., 45, 503

\bibitem[{L\"owdin(1955)}]{Lowdin:1955}
L\"owdin, P.-O. 1955, Phys. Rev., 97, 1474

\bibitem[{Mardirossian \& Head-Gordon(2016)}]{Mardirossian_2016}
Mardirossian, N. \& Head-Gordon, M. 2016, J. Chem. Theory Comput., 12, 4303

\bibitem[{Matta \& Hernández-Trujillo(2003)}]{Cherif_2003}
Matta, C.~F. \& Hernández-Trujillo, J. 2003, J. Phys. Chem. A, 107, 7496

\bibitem[{McConnell \& K\"astner(2017)}]{mcc17a}
McConnell, S. \& K\"astner, J. 2017, J. Comput. Chem., 38, 2570

\bibitem[{Meisner {et~al.}(2017)Meisner, Lamberts, \& Kästner}]{Meisner_2017}
Meisner, J., Lamberts, T., \& Kästner, J. 2017, ACS Earth and Space Chem., 1, 399

\bibitem[{Metropolis \& Ulam(1949)}]{MonteCarlo_1949}
Metropolis, N. \& Ulam, S. 1949, JASA, 44, 335

\bibitem[{Metz {et~al.}(2014)Metz, K\"astner, Sokol, Keal, \& Sherwood}]{met14}
Metz, S., K\"astner, J., Sokol, A.~A., Keal, T.~W., \& Sherwood, P. 2014, WIREs Comput. Mol. Sci., 4, 101

\bibitem[{Miller {et~al.}(1983)Miller, Schwartz, \& Tromp}]{Miller:1983}
Miller, W.~H., Schwartz, S.~D., \& Tromp, J.~W. 1983, J. Phys. Chem., 79, 4889

\bibitem[{{Montillaud, J.} {et~al.}(2013){Montillaud, J.}, {Joblin, C.}, \& {Toublanc, D.}}]{Montillaud_2013}
{Montillaud, J.}, {Joblin, C.}, \& {Toublanc, D.} 2013, A\&A, 552, A15

\bibitem[{Mulliken(1955)}]{Mulliken_1955}
Mulliken, R.~S. 1955, J. Chem. Phys, 23, 1833

\bibitem[{Neese {et~al.}(2020)Neese, Wennmohs, Becker, \& Riplinger}]{Neese_2020}
Neese, F., Wennmohs, F., Becker, U., \& Riplinger, C. 2020, J. Chem. Phys., 152, 224108

\bibitem[{Rapacioli {et~al.}(2005)Rapacioli, Calvo, Spiegelman, Joblin, \& Wales}]{Rapacioli_2005}
Rapacioli, M., Calvo, F., Spiegelman, F., Joblin, C., \& Wales, D.~J. 2005, J. Phys. Chem. A, 109, 2487

\bibitem[{{Rapacioli, M.} {et~al.}(2006){Rapacioli, M.}, {Calvo, F.}, {Joblin, C.}, {Parneix, P.}, {Toublanc, D.}, \& {Spiegelman, F.}}]{Rapacioli_2006}
{Rapacioli, M.}, {Calvo, F.}, {Joblin, C.}, {et~al.} 2006, A\&A, 460, 519

\bibitem[{Rauls \& Hornekær(2008)}]{Rauls_2008}
Rauls, E. \& Hornekær, L. 2008, ApJ, 679, 531

\bibitem[{Rommel {et~al.}(2011)Rommel, Goumans, \& Kästner}]{Rommel:Instanton:2011}
Rommel, J.~B., Goumans, T. P.~M., \& Kästner, J. 2011, J. Chem. Theory Comput., 7, 690

\bibitem[{Rommel \& Kästner(2011)}]{Rommel:2011}
Rommel, J.~B. \& Kästner, J. 2011, J. Chem. Phys., 134, 184107

\bibitem[{Schneiker {et~al.}(2024)Schneiker, Góbi, Ragupathy, Keresztes, Bazsó, \& Tarczay}]{Indene_2024}
Schneiker, A., Góbi, S., Ragupathy, G., {et~al.} 2024, J. Chem. Phys., 160, 214303

\bibitem[{Sephton \& Botta(2005)}]{Sephton_Botta_2005}
Sephton, M.~A. \& Botta, O. 2005, Int. J. Astrobiol., 4, 269–276

\bibitem[{{Tang, Z.} {et~al.}(2022){Tang, Z.}, {Simonsen, F. D. S.}, {Jaganathan, R.}, {Palotás, J.}, {Oomens, J.}, {Hornekær, L.}, \& {Hammer, B.}}]{Zeyuan_2022}
{Tang, Z.}, {Simonsen, F. D. S.}, {Jaganathan, R.}, {et~al.} 2022, A\&A, 663, A150

\bibitem[{Tielens(2008)}]{Tielens_2008}
Tielens, A. G. G.~M. 2008, Annu. Rev. Astron. Astrophys., 46, 289

\bibitem[{Tielens(2013)}]{Tielens_2013}
Tielens, A. G. G.~M. 2013, Rev. Mod. Phys., 85, 1021

\bibitem[{Wakelam {et~al.}(2017)Wakelam, Bron, Cazaux, Dulieu, Gry, Guillard, Habart, Hornekær, Morisset, Nyman, Pirronello, Price, Valdivia, Vidali, \& Watanabe}]{Wakelam2017}
Wakelam, V., Bron, E., Cazaux, S., {et~al.} 2017, Mol. Astrophys., 9, 1

\bibitem[{Wakelam {et~al.}(2012)Wakelam, Herbst, Loison, Smith, Chandrasekaran, Pavone, Adams, Bacchus-Montabonel, Bergeat, Béroff, Bierbaum, Chabot, Dalgarno, van Dishoeck, Faure, Geppert, Gerlich, Galli, Hébrard, Hersant, Hickson, Honvault, Klippenstein, Picard, Nyman, Pernot, Schlemmer, Selsis, Sims, Talbi, Tennyson, Troe, Wester, \& Wiesenfeld}]{Wakelam_2012}
Wakelam, V., Herbst, E., Loison, J.-C., {et~al.} 2012, The Astrophysical Journal Supplement Series, 199, 21

\bibitem[{Yamijala {et~al.}(2019)Yamijala, Ali, \& Wong}]{TZP_des_2019}
Yamijala, S. S. R. K.~C., Ali, Z.~A., \& Wong, B.~M. 2019, J. Phys. Chem. C, 123, 25113

\bibitem[{Zhao \& Truhlar(2008)}]{Zhao2008}
Zhao, Y. \& Truhlar, D.~G. 2008, Theor. Chem. Acc., 120, 215

\bibitem[{Zheng \& Truhlar(2010)}]{Zheng_2010}
Zheng, J. \& Truhlar, D.~G. 2010, Phys. Chem. Chem. Phys., 12, 7782

\end{thebibliography}

\appendix

\setcounter{section}{1}  

\renewcommand{\thesection}{\Alph{section}} 
\renewcommand{\thefigure}{\thesection.\arabic{figure}} 
\renewcommand{\thetable}{\thesection.\arabic{table}}   

\setcounter{figure}{0} 
\setcounter{table}{0}  

\section*{Appendix A: Indene hydrogenation}
\addcontentsline{toc}{section}{Appendix \thesection: Indene hydrogenation}
\label{sec:hydrogenation_SI}
We report the binding energies and barrier energies of the odd hydrogenation sequence of indene in Tables~\ref{table:Indene_1H}, \ref{table:Indene_3H}, \ref{table:Indene_5H},{~and \ref{table:Indene_7H}}, while the odd hydrogenated structures of indene are shown in Figure~\ref{Fig:Indene_sequence}. The even hydrogenation of indene was considered by locating the unpaired electron through spin population analysis, and their binding energy values are reported in Table \ref{table:Indene_evenH}. 

\begin{table}[h]
\caption{Binding energies ($E_\text{b}$) and energy barriers ($E_\text{act}$) for the 1st hydrogenation of indene.}            
\label{table:Indene_1H}     
\centering                         
\begin{tabular}{c c c}       
\hline\hline                
C site & $E_\text{b}$ (eV) & $E_\text{act}$ (eV) \\ 
\hline                        
   1 & 1.67  & 0.13 \\    
   2 & 1.32  & 0.22 \\
   3 & 0.47  & 0.40 \\
   4 & 0.95 & 0.25 \\
   5 & 0.80 & 0.29 \\
   6 & 0.98 & 0.26 \\
   7 & 0.83 & 0.26 \\
   8 & 0.78 & 0.28 \\
\hline       
\end{tabular}
\tablefoot{C site labels are associated with the values reported in Fig. \ref{Fig:Indene}.}
\end{table}

\begin{table}[h]
\caption{Binding energies ($E_\text{b}$) and energy barriers ($E_\text{act}$) for the 3rd hydrogenation of indene.}             
\label{table:Indene_3H}      
\centering                         
\begin{tabular}{c c c}       
\hline\hline                
C site & $E_\text{b}$ (eV) & $E_\text{act}$ (eV) \\     
\hline                        
   3 & 0.88  & 0.24 \\      
   5 & 0.72  & 0.26 \\
   7 & 0.83  & 0.27 \\
\hline                                
\end{tabular}
\tablefoot{C site labels are associated with the values reported in Fig. \ref{Fig:Indene}.}
\end{table}

\begin{table}[h]
\caption{Binding energies ($E_\text{b}$) and energy barriers ($E_\text{act}$) for the 5th hydrogenation of indene.}            
\label{table:Indene_5H}    
\centering                         
\begin{tabular}{c c c}        
\hline\hline                
C site & $E_\text{b}$ (eV) & $E_\text{act}$ (eV) \\    
\hline                      
   5 & 1.26  & 0.13 \\      
   {7} & 1.33  & 0.18 \\
\hline                                   
\end{tabular}
\tablefoot{C site labels are associated with the values reported in Fig. \ref{Fig:Indene}.}
\end{table}

\begin{table}[h]
\caption{Binding energies ($E_\text{b}$) and energy barriers ($E_\text{act}$) for the 7th hydrogenation of indene.}            
\label{table:Indene_7H}      
\centering                         
\begin{tabular}{c c c}       
\hline\hline                 
C site & $E_\text{b}$ (eV) & $E_\text{act}$ (eV) \\ 
\hline                       
   7 & 1.42  & 0.14 \\      
\hline                             
\end{tabular}
\tablefoot{C site labels are associated with the values reported in Fig. \ref{Fig:Indene}.}
\end{table}

\begin{table}[h]
\caption{Binding Energies ($E_{b}$) for the even hydrogenation (nH) of indene.}            
\label{table:Indene_evenH}      
\centering                          
\begin{tabular}{c c}        
\hline\hline                 
 nH & $E_{b}$ (eV) \\    
\hline                        
   2H & 3.71  \\     
   4H & 3.24  \\
   6H & 4.00  \\
   8H & 4.14  \\
\hline                                   
\end{tabular}
\end{table}

\stepcounter{section} 
\renewcommand{\thesection}{\Alph{section}}
\renewcommand{\thefigure}{\thesection.\arabic{figure}}
\renewcommand{\thetable}{\thesection.\arabic{table}}
\setcounter{figure}{0}
\setcounter{table}{0}

\section*{Appendix B: Rate fitting}
\addcontentsline{toc}{section}{Appendix \thesection: Rate fitting}
\label{sec:Rate_fitting_SI}
{We report here the values of rate constants computed at the DFT level, corrected for both Eckart and Bell tunneling, along with the corresponding rate constants obtained using equation \ref{eq5} by applying the fitted parameters obtained for each hydrogenation step in indene, as shown in Tables \ref{tab:fitted_bell_1H} to \ref{tab:fitted_eckart_7H}.}

{The initial guess bounds for the Bell fitting are (1x10$^{-16}$, 1x10$^{-9}$) {cm$^3$ s$^{-1}$}, (0.01, 5), (1, 4000) {K}, and (10, 1000) {K} for $\alpha$, $\beta$, $\gamma$, and T$_0$, respectively, while for the Eckart fitting, the bounds are (1x10$^{-14}$, 1x10$^{-9}$) {cm$^3$ s$^{-1}$}, (1, 5), (10, 1000) {K}, and (10, 100) {K} for $\alpha$, $\beta$, $\gamma$, and T$_0$, respectively.}
The bounds are needed solely to constrain the initial guess formation for obtaining the best fit. 

We also report rate constants based on the instanton \citep{Miller:1983,Rommel:Instanton:2011} and asymmetric Eckart \citep{Asym_Eckart} rate constants, together with the corresponding fitted parameters (Tables \ref{tab:AsymEckart_fit_1H} - \ref{tab:AsymEckart_fit_7H}). The computed asymmetric Eckart rate constants result in negligible differences at higher temperatures and differences of less than one order of magnitude at lower temperatures compared to the symmetric Eckart and instanton rate constants. Furthermore, instanton calculations diverge by less than an order of magnitude with respect to symmetric and asymmetric Eckart. It should be noted that for instanton rate constants, we report values only up to 80-50 K due to difficulties in converging the Feynman pathways below this temperature. Both asymmetric Eckart and instanton rate constants are computed using DL-FIND in CHEMSHELL \citep{DL-FIND,met14}. The Feynman pathway for instanton calculations has been optimized using the Newton-Raphson optimization algorithm \citep{Rommel:2011,Rommel:Instanton:2011}. The Feynman pathway was considered to have converged when the gradient fell below 9 x 10$^{-8}$ a.u./bohr. We used between 15 and 22 images, with a larger number of images required to converge the pathway at lower temperatures. After optimization, the Hessian was computed for all images to determine the instanton rate. Input and output files for the asymmetric Eckart and instanton calculations are provided in the Zenodo repository \citep{Campisi_2024}.
The fitted parameters for the asymmetric Eckart have been determined with our code as described in Section \ref{Implementation}, using as bounds (1 × 10$^{-14}$, 1 × 10$^{-10}$) cm$^3$ s$^{-1}$, (1, 3), (1, 2000) K and (10, 100) K, for $\alpha$, $\beta$, $\gamma$ and T$_0$, respectively. For instantons, we employed the same bounds as those of the asymmetric Eckart for the addition of the first, third, and fifth hydrogenations of indene. For the seventh addition of hydrogen to indene, the initial guess bounds to determine the rate constants are (1 × 10$^{-14}$, 1 × 10$^{-10}$) {cm$^3$ s$^{-1}$}, (1, 3), (1, 2000) K and (50, 200) K, for $\alpha$, $\beta$, $\gamma$, and T$_0$, respectively.

\begin{table}
    \centering
    \caption{Bimolecular Bell rate constants for the 1st hydrogenation of indene.}
    \begin{tabular}{lcc}
        \hline
        \hline
        Temperature & Bell &  Fitted\\
        \hline
        10000 & 5.63 x 10$^{-10}$ & 5.49 x 10$^{-10}$ \\
        282.10 & 8.66 x 10$^{-14}$ & 1.05 x 10$^{-13}$ \\
        143.07 & 3.20 x 10$^{-15}$ & 2.81 x 10$^{-15}$ \\
        95.84 & 1.10 x 10$^{-15}$ & 9.71 x 10$^{-16}$ \\
        72.05 & 8.40 x 10$^{-16}$ & 7.44 x 10$^{-16}$ \\
        57.72 & 7.79 x 10$^{-16}$ & 7.22 x 10$^{-16}$ \\
        48.15 & 7.67 x 10$^{-16}$ & 7.49 x 10$^{-16}$ \\
        41.30 & 7.73 x 10$^{-16}$ & 7.91 x 10$^{-16}$ \\
        36.16 & 7.87 x 10$^{-16}$ & 8.34 x 10$^{-16}$ \\
        32.15 & 8.05 x 10$^{-16}$ & 8.75 x 10$^{-16}$ \\
        28.95 & 8.25 x 10$^{-16}$ & 9.11 x 10$^{-16}$ \\
        26.32 & 8.46 x 10$^{-16}$ & 9.43 x 10$^{-16}$ \\
        24.13 & 8.68 x 10$^{-16}$ & 9.69 x 10$^{-16}$ \\
        22.28 & 8.90 x 10$^{-16}$ & 9.92 x 10$^{-16}$ \\
        20.69 & 9.12 x 10$^{-16}$ & 1.01 x 10$^{-15}$ \\
        19.32 & 9.34 x 10$^{-16}$ & 1.03 x 10$^{-15}$ \\
        18.11 & 9.55 x 10$^{-16}$ & 1.04 x 10$^{-15}$ \\
        17.05 & 9.77 x 10$^{-16}$ & 1.05 x 10$^{-15}$ \\
        16.10 & 9.98 x 10$^{-16}$ & 1.06 x 10$^{-15}$ \\
        15.26 & 1.02 x 10$^{-15}$ & 1.06 x 10$^{-15}$ \\
        14.49 & 1.04 x 10$^{-15}$ & 1.07 x 10$^{-15}$ \\
        13.80 & 1.06 x 10$^{-15}$ & 1.07 x 10$^{-15}$ \\
        13.18 & 1.08 x 10$^{-15}$ & 1.07 x 10$^{-15}$ \\
        12.61 & 1.10 x 10$^{-15}$ & 1.07 x 10$^{-15}$ \\
        12.08 & 1.12 x 10$^{-15}$ & 1.07 x 10$^{-15}$ \\
        11.60 & 1.14 x 10$^{-15}$ & 1.07 x 10$^{-15}$ \\
        11.15 & 1.16 x 10$^{-15}$ & 1.07 x 10$^{-15}$ \\
        10.74 & 1.17 x 10$^{-15}$ & 1.06 x 10$^{-15}$ \\
        10.36 & 1.19 x 10$^{-15}$ & 1.06 x 10$^{-15}$ \\
        10 & 1.21 x 10$^{-15}$ & 1.06 x 10$^{-15}$ \\
        \hline
        \hline
    \end{tabular}
    \label{tab:fitted_bell_1H}
    \tablefoot{Temperature, in kelvin (K), DFT-computed "Bell" rate constants, in cm$^3$ s$^{-1}$, and "fitted" Bell rate constants, in cm$^3$ s$^{-1}$, obtained by using the fitted parameters ($\alpha$: 9.857 x 10$^{-11}$  cm$^3$ s$^{-1}$, $\beta$: 0.538 , $\gamma$: 1652.862 K, $T_{0}$: 180.897 K, RMS: 0.038, R$^{2}$: 0.999) in equation \ref{eq5}.}
\end{table}

\begin{table}
    \centering
    \caption{Bimolecular Eckart rate constants for the 1st hydrogenation of indene.}
    \begin{tabular}{lcc}
        \hline
        \hline
        Temperature & Eckart &  Fitted\\
        \hline
        10000 & 5.64 x 10$^{-10}$ & 6.79 x 10$^{-10}$ \\
        282.10 & 8.73 x 10$^{-14}$ & 6.36 x 10$^{-14}$ \\
        143.07 & 1.67 x 10$^{-15}$ & 1.80 x 10$^{-15}$ \\
        95.84 & 1.13 x 10$^{-16}$ & 1.42 x 10$^{-16}$ \\
        72.05 & 2.16 x 10$^{-17}$ & 2.52 x 10$^{-17}$ \\
        57.72 & 7.78 x 10$^{-18}$ & 7.92 x 10$^{-18}$ \\
        48.15 & 3.92 x 10$^{-18}$ & 3.64 x 10$^{-18}$ \\
        41.30 & 2.39 x 10$^{-18}$ & 2.13 x 10$^{-18}$ \\
        36.16 & 1.65 x 10$^{-18}$ & 1.46 x 10$^{-18}$ \\
        32.15 & 1.23 x 10$^{-18}$ & 1.10 x 10$^{-18}$ \\
        28.95 & 9.71 x 10$^{-19}$ & 8.83 x 10$^{-19}$ \\
        26.32 & 7.98 x 10$^{-19}$ & 7.43 x 10$^{-19}$ \\
        24.13 & 6.76 x 10$^{-19}$ & 6.44 x 10$^{-19}$ \\
        22.28 & 5.86 x 10$^{-19}$ & 5.70 x 10$^{-19}$ \\
        20.69 & 5.17 x 10$^{-19}$ & 5.13 x 10$^{-19}$ \\
        19.32 & 4.63 x 10$^{-19}$ & 4.67 x 10$^{-19}$ \\
        18.11 & 4.20 x 10$^{-19}$ & 4.29 x 10$^{-19}$ \\
        17.05 & 3.85 x 10$^{-19}$ & 3.97 x 10$^{-19}$ \\
        16.10 & 3.56 x 10$^{-19}$ & 3.70 x 10$^{-19}$ \\
        15.26 & 3.31 x 10$^{-19}$ & 3.46 x 10$^{-19}$ \\
        14.49 & 3.10 x 10$^{-19}$ & 3.24 x 10$^{-19}$ \\
        13.80 & 2.92 x 10$^{-19}$ & 3.05 x 10$^{-19}$ \\
        13.18 & 2.76 x 10$^{-19}$ & 2.88 x 10$^{-19}$ \\
        12.61 & 2.62 x 10$^{-19}$ & 2.72 x 10$^{-19}$ \\
        12.08 & 2.49 x 10$^{-19}$ & 2.58 x 10$^{-19}$ \\
        11.60 & 2.38 x 10$^{-19}$ & 2.45 x 10$^{-19}$ \\
        11.15 & 2.28 x 10$^{-19}$ & 2.33 x 10$^{-19}$ \\
        10.74 & 2.19 x 10$^{-19}$ & 2.22 x 10$^{-19}$ \\
        10.36 & 2.11 x 10$^{-19}$ & 2.12 x 10$^{-19}$ \\
        10 & 2.04 x 10$^{-19}$ & 2.03 x 10$^{-19}$ \\
        \hline
        \hline
    \end{tabular}
    \label{tab:fitted_eckart_1H}
    \tablefoot{Temperature, in kelvin (K), DFT-computed "Eckart" rate constants, in cm$^3$ s$^{-1}$, and "fitted" Eckart rate constants, in cm$^3$ s$^{-1}$, obtained by using the fitted parameters ($\alpha$: 6.437 x 10$^{-13}$ cm$^3$ s$^{-1}$, $\beta$: 2.000, $\gamma$: 524.747 K, $T_{0}$: 71.797 K, RMS: 0.043, R$^2$: 0.999) in equation \ref{eq5}.}
\end{table}

\begin{table}
    \centering
    \caption{Bimolecular Bell rate constants for the 3rd hydrogenation of indene.}
    \begin{tabular}{lcc}
        \hline
        \hline
        Temperature & Bell &  Fitted\\
        \hline
        10000 & 5.26 x 10$^{-10}$ & 3.59 x 10$^{-10}$ \\
        282.10 & 1.91 x 10$^{-15}$ & 2.00 x 10$^{-15}$ \\
        143.07 & 3.52 x 10$^{-17}$ & 3.83 x 10$^{-17}$ \\
        95.84 & 2.06 x 10$^{-17}$ & 1.66 x 10$^{-17}$ \\
        72.05 & 1.84 x 10$^{-17}$ & 1.50 x 10$^{-17}$ \\
        57.72 & 1.81 x 10$^{-17}$ & 1.59 x 10$^{-17}$ \\
        48.15 & 1.83 x 10$^{-17}$ & 1.74 x 10$^{-17}$ \\
        41.30 & 1.88 x 10$^{-17}$ & 1.90 x 10$^{-17}$ \\
        36.16 & 1.94 x 10$^{-17}$ & 2.04 x 10$^{-17}$ \\
        32.15 & 2.00 x 10$^{-17}$ & 2.16 x 10$^{-17}$ \\
        28.95 & 2.07 x 10$^{-17}$ & 2.27 x 10$^{-17}$ \\
        26.32 & 2.13 x 10$^{-17}$ & 2.36 x 10$^{-17}$ \\
        24.13 & 2.20 x 10$^{-17}$ & 2.43 x 10$^{-17}$ \\
        22.28 & 2.26 x 10$^{-17}$ & 2.49 x 10$^{-17}$ \\
        20.69 & 2.33 x 10$^{-17}$ & 2.53 x 10$^{-17}$ \\
        19.32 & 2.39 x 10$^{-17}$ & 2.57 x 10$^{-17}$ \\
        18.11 & 2.45 x 10$^{-17}$ & 2.59 x 10$^{-17}$ \\
        17.05 & 2.51 x 10$^{-17}$ & 2.61 x 10$^{-17}$ \\
        16.10 & 2.57 x 10$^{-17}$ & 2.63 x 10$^{-17}$ \\
        15.26 & 2.63 x 10$^{-17}$ & 2.63 x 10$^{-17}$ \\
        14.49 & 2.69 x 10$^{-17}$ & 2.64 x 10$^{-17}$ \\
        13.80 & 2.74 x 10$^{-17}$ & 2.64 x 10$^{-17}$ \\
        13.18 & 2.80 x 10$^{-17}$ & 2.63 x 10$^{-17}$ \\
        12.61 & 2.85 x 10$^{-17}$ & 2.63 x 10$^{-17}$ \\
        12.08 & 2.90 x 10$^{-17}$ & 2.62 x 10$^{-17}$ \\
        11.60 & 2.96 x 10$^{-17}$ & 2.61 x 10$^{-17}$ \\
        11.15 & 3.01 x 10$^{-17}$ & 2.60 x 10$^{-17}$ \\
        10.74 & 3.06 x 10$^{-17}$ & 2.59 x 10$^{-17}$ \\
        10.36 & 3.11 x 10$^{-17}$ & 2.57 x 10$^{-17}$ \\
        10 & 3.16 x 10$^{-17}$ & 2.56 x 10$^{-17}$ \\
        \hline
        \hline
    \end{tabular}
    \label{tab:fitted_bell_3H}
    \tablefoot{Temperature, in kelvin (K), DFT-computed "Bell" rate constants, in cm$^3$ s$^{-1}$ and "fitted" Bell rate constants, in cm$^3$ s$^{-1}$, obtained by using the fitted parameters ($\alpha$: 4.732 x 10$^{-11}$ cm$^3$ s$^{-1}$, $\beta$: 0.652, $\gamma$: 2549.480 K, $T_{0}$: 217.898 K, RMS: 0.055, R$^{2}$: 0.998) in equation \ref{eq5}.}
\end{table}

\begin{table}
    \centering
    \caption{Bimolecular Eckart rate constants for the 3rd hydrogenation of indene.}
    \begin{tabular}{lcc}
        \hline
        \hline
        Temperature & Eckart &  Fitted\\
        \hline
        10000 & 5.27 x 10$^{-10}$ & 3.54 x 10$^{-10}$ \\
        282.10 & 1.73 x 10$^{-15}$ & 1.07 x 10$^{-15}$ \\
        143.07 & 2.66 x 10$^{-18}$ & 4.50 x 10$^{-18}$ \\
        95.84 & 8.98 x 10$^{-20}$ & 1.35 x 10$^{-19}$ \\
        72.05 & 1.57 x 10$^{-20}$ & 1.65 x 10$^{-20}$ \\
        57.72 & 5.53 x 10$^{-21}$ & 4.69 x 10$^{-21}$ \\
        48.15 & 2.75 x 10$^{-21}$ & 2.17 x 10$^{-21}$ \\
        41.30 & 1.67 x 10$^{-21}$ & 1.31 x 10$^{-21}$ \\
        36.16 & 1.14 x 10$^{-21}$ & 9.30 x 10$^{-22}$ \\
        32.15 & 8.48 x 10$^{-22}$ & 7.24 x 10$^{-22}$ \\
        28.95 & 6.67 x 10$^{-22}$ & 5.98 x 10$^{-22}$ \\
        26.32 & 5.47 x 10$^{-22}$ & 5.12 x 10$^{-22}$ \\
        24.13 & 4.62 x 10$^{-22}$ & 4.50 x 10$^{-22}$ \\
        22.28 & 4.00 x 10$^{-22}$ & 4.02 x 10$^{-22}$ \\
        20.69 & 3.53 x 10$^{-22}$ & 3.64 x 10$^{-22}$ \\
        19.32 & 3.16 x 10$^{-22}$ & 3.32 x 10$^{-22}$ \\
        18.11 & 2.86 x 10$^{-22}$ & 3.05 x 10$^{-22}$ \\
        17.05 & 2.62 x 10$^{-22}$ & 2.82 x 10$^{-22}$ \\
        16.10 & 2.42 x 10$^{-22}$ & 2.62 x 10$^{-22}$ \\
        15.26 & 2.25 x 10$^{-22}$ & 2.44 x 10$^{-22}$ \\
        14.49 & 2.11 x 10$^{-22}$ & 2.28 x 10$^{-22}$ \\
        13.80 & 1.98 x 10$^{-22}$ & 2.13 x 10$^{-22}$ \\
        13.18 & 1.87 x 10$^{-22}$ & 2.00 x 10$^{-22}$ \\
        12.61 & 1.78 x 10$^{-22}$ & 1.88 x 10$^{-22}$ \\
        12.08 & 1.69 x 10$^{-22}$ & 1.77 x 10$^{-22}$ \\
        11.60 & 1.62 x 10$^{-22}$ & 1.67 x 10$^{-22}$ \\
        11.15 & 1.55 x 10$^{-22}$ & 1.58 x 10$^{-22}$ \\
        10.74 & 1.49 x 10$^{-22}$ & 1.49 x 10$^{-22}$ \\
        10.36 & 1.43 x 10$^{-22}$ & 1.41 x 10$^{-22}$ \\
        10 & 1.38 x 10$^{-22}$ & 1.34 x 10$^{-22}$ \\
        \hline
        \hline
    \end{tabular}
    \label{tab:fitted_eckart_3H}
    \tablefoot{Temperature, in kelvin (K), DFT-computed "Eckart" rate constants, in cm$^3$ s$^{-1}$, and "fitted" Eckart rate constants, in cm$^3$ s$^{-1}$, obtained by using the fitted parameters ($\alpha$: 7.755 x 10$^{-14}$ cm$^3$ s$^{-1}$, $\beta$: 2.431, $\gamma$: 974.693 K, $T_{0}$: 89.832 K, RMS: 0.083, R$^{2}$: 0.999) in equation \ref{eq5}.}
\end{table}

\begin{table}
    \centering
    \caption{Bimolecular Bell rate constants for the 5th hydrogenation of indene.}
    \begin{tabular}{lcc}
        \hline
        \hline
        Temperature & Bell & Fitted \\
        \hline
        10000 & 1.41 x 10$^{-10}$ & 1.08 x 10$^{-10}$ \\
        282.10 & 6.19 x 10$^{-14}$ & 6.60 x 10$^{-14}$ \\
        143.07 & 6.44 x 10$^{-15}$ & 5.14 x 10$^{-15}$ \\
        95.84 & 3.72 x 10$^{-15}$ & 3.07 x 10$^{-15}$ \\
        72.05 & 3.40 x 10$^{-15}$ & 2.96 x 10$^{-15}$ \\
        57.72 & 3.42 x 10$^{-15}$ & 3.17 x 10$^{-15}$ \\
        48.15 & 3.54 x 10$^{-15}$ & 3.46 x 10$^{-15}$ \\
        41.30 & 3.68 x 10$^{-15}$ & 3.74 x 10$^{-15}$ \\
        36.16 & 3.82 x 10$^{-15}$ & 4.01 x 10$^{-15}$ \\
        32.15 & 3.97 x 10$^{-15}$ & 4.24 x 10$^{-15}$ \\
        28.95 & 4.11 x 10$^{-15}$ & 4.45 x 10$^{-15}$ \\
        26.32 & 4.24 x 10$^{-15}$ & 4.63 x 10$^{-15}$ \\
        24.13 & 4.37 x 10$^{-15}$ & 4.79 x 10$^{-15}$ \\
        22.28 & 4.50 x 10$^{-15}$ & 4.93 x 10$^{-15}$ \\
        20.69 & 4.63 x 10$^{-15}$ & 5.04 x 10$^{-15}$ \\
        19.32 & 4.75 x 10$^{-15}$ & 5.15 x 10$^{-15}$ \\
        18.11 & 4.87 x 10$^{-15}$ & 5.23 x 10$^{-15}$ \\
        17.05 & 4.98 x 10$^{-15}$ & 5.31 x 10$^{-15}$ \\
        16.10 & 5.10 x 10$^{-15}$ & 5.38 x 10$^{-15}$ \\
        15.26 & 5.21 x 10$^{-15}$ & 5.43 x 10$^{-15}$ \\
        14.49 & 5.32 x 10$^{-15}$ & 5.48 x 10$^{-15}$ \\
        13.80 & 5.42 x 10$^{-15}$ & 5.52 x 10$^{-15}$ \\
        13.18 & 5.53 x 10$^{-15}$ & 5.56 x 10$^{-15}$ \\
        12.61 & 5.63 x 10$^{-15}$ & 5.59 x 10$^{-15}$ \\
        12.08 & 5.74 x 10$^{-15}$ & 5.62 x 10$^{-15}$ \\
        11.60 & 5.84 x 10$^{-15}$ & 5.64 x 10$^{-15}$ \\
        11.15 & 5.93 x 10$^{-15}$ & 5.66 x 10$^{-15}$ \\
        10.74 & 6.03 x 10$^{-15}$ & 5.67 x 10$^{-15}$ \\
        10.36 & 6.13 x 10$^{-15}$ & 5.68 x 10$^{-15}$ \\
        10 & 6.22 x 10$^{-15}$ & 5.69 x 10$^{-15}$ \\
        \hline
        \hline
    \end{tabular}
    \label{tab:fitted_bell_5H}
    \tablefoot{Temperature, in kelvin (K), DFT-computed "Bell" rate constants, in cm$^3$ s$^{-1}$, and "fitted" Bell rate constants, in cm$^3$ s$^{-1}$, obtained by using the fitted parameters ($\alpha$: 4.778 x 10$^{-11}$ cm$^3$ s$^{-1}$, $\beta$: 0.280, $\gamma$: 1661.515 K, $T_{0}$: 214.678 K, RMS: 0.042, R$^{2}$: 0.997) in equation \ref{eq5}.}
\end{table}

\begin{table}
    \centering
    \caption{Bimolecular Eckart rate constants for the 5th hydrogenation of indene.}
    \begin{tabular}{lcc}
        \hline
        \hline
        Temperature & Eckart & Fitted \\
        \hline
        10000 & 1.42 x 10$^{-10}$ & 1.97 x 10$^{-10}$ \\
        282.10 & 6.10 x 10$^{-14}$ & 5.69 x 10$^{-14}$ \\
        143.07 & 2.70 x 10$^{-15}$ & 2.61 x 10$^{-15}$ \\
        95.84 & 3.76 x 10$^{-16}$ & 3.66 x 10$^{-16}$ \\
        72.05 & 1.19 x 10$^{-16}$ & 1.10 x 10$^{-16}$ \\
        57.72 & 5.84 x 10$^{-17}$ & 5.19 x 10$^{-17}$ \\
        48.15 & 3.61 x 10$^{-17}$ & 3.20 x 10$^{-17}$ \\
        41.30 & 2.55 x 10$^{-17}$ & 2.30 x 10$^{-17}$ \\
        36.16 & 1.95 x 10$^{-17}$ & 1.81 x 10$^{-17}$ \\
        32.15 & 1.57 x 10$^{-17}$ & 1.50 x 10$^{-17}$ \\
        28.95 & 1.32 x 10$^{-17}$ & 1.29 x 10$^{-17}$ \\
        26.32 & 1.13 x 10$^{-17}$ & 1.14 x 10$^{-17}$ \\
        24.13 & 9.98 x 10$^{-18}$ & 1.02 x 10$^{-17}$ \\
        22.28 & 8.93 x 10$^{-18}$ & 9.31 x 10$^{-18}$ \\
        20.69 & 8.10 x 10$^{-18}$ & 8.53 x 10$^{-18}$ \\
        19.32 & 7.43 x 10$^{-18}$ & 7.88 x 10$^{-18}$ \\
        18.11 & 6.87 x 10$^{-18}$ & 7.31 x 10$^{-18}$ \\
        17.05 & 6.40 x 10$^{-18}$ & 6.82 x 10$^{-18}$ \\
        16.10 & 6.00 x 10$^{-18}$ & 6.38 x 10$^{-18}$ \\
        15.26 & 5.66 x 10$^{-18}$ & 6.00 x 10$^{-18}$ \\
        14.49 & 5.36 x 10$^{-18}$ & 5.65 x 10$^{-18}$ \\
        13.80 & 5.10 x 10$^{-18}$ & 5.33 x 10$^{-18}$ \\
        13.18 & 4.87 x 10$^{-18}$ & 5.05 x 10$^{-18}$ \\
        12.61 & 4.66 x 10$^{-18}$ & 4.79 x 10$^{-18}$ \\
        12.08 & 4.47 x 10$^{-18}$ & 4.55 x 10$^{-18}$ \\
        11.60 & 4.31 x 10$^{-18}$ & 4.33 x 10$^{-18}$ \\
        11.15 & 4.15 x 10$^{-18}$ & 4.12 x 10$^{-18}$ \\
        10.74 & 4.01 x 10$^{-18}$ & 3.94 x 10$^{-18}$ \\
        10.36 & 3.89 x 10$^{-18}$ & 3.76 x 10$^{-18}$ \\
        10 & 3.77 x 10$^{-18}$ & 3.60 x 10$^{-18}$ \\
        \hline
        \hline
    \end{tabular}
\label{tab:fitted_eckart_5H}
\tablefoot{Temperature, in kelvin (K), DFT-computed "Eckart" rate constants, in cm$^3$ s$^{-1}$, and "fitted" Eckart rate constants, in cm$^3$ s$^{-1}$, obtained by using the fitted parameters ($\alpha$: 4.99 x 10$^{-13}$ cm$^3$ s$^{-1}$, $\beta$: 1.718, $\gamma$: 487.141 K, $T_{0}$: 89.230 K, RMS: 0.035, R$^{2}$: 0.999) in equation \ref{eq5}.}
\end{table}

\begin{table}
    \centering
    \caption{Bimolecular Bell rate constants for the 7th hydrogenation of indene.}
    \begin{tabular}{lcc}
        \hline
        \hline
        Temperature & Bell & Fitted \\
        \hline
        10000 & 3.37 x 10$^{-10}$ & 2.92 x 10$^{-10}$ \\
        282.10 & 5.93 x 10$^{-14}$ & 5.95 x 10$^{-14}$ \\
        143.07 & 3.00 x 10$^{-15}$ & 2.42 x 10$^{-15}$ \\
        95.84 & 1.34 x 10$^{-15}$ & 1.09 x 10$^{-15}$ \\
        72.05 & 1.11 x 10$^{-15}$ & 9.39 x 10$^{-16}$ \\
        57.72 & 1.06 x 10$^{-15}$ & 9.58 x 10$^{-16}$ \\
        48.15 & 1.06 x 10$^{-15}$ & 1.02 x 10$^{-15}$ \\
        41.30 & 1.08 x 10$^{-15}$ & 1.09 x 10$^{-15}$ \\
        36.16 & 1.10 x 10$^{-15}$ & 1.15 x 10$^{-15}$ \\
        32.15 & 1.13 x 10$^{-15}$ & 1.21 x 10$^{-15}$ \\
        28.95 & 1.16 x 10$^{-15}$ & 1.26 x 10$^{-15}$ \\
        26.32 & 1.19 x 10$^{-15}$ & 1.30 x 10$^{-15}$ \\
        24.13 & 1.23 x 10$^{-15}$ & 1.34 x 10$^{-15}$ \\
        22.28 & 1.26 x 10$^{-15}$ & 1.37 x 10$^{-15}$ \\
        20.69 & 1.29 x 10$^{-15}$ & 1.39 x 10$^{-15}$ \\
        19.32 & 1.32 x 10$^{-15}$ & 1.41 x 10$^{-15}$ \\
        18.11 & 1.35 x 10$^{-15}$ & 1.43 x 10$^{-15}$ \\
        17.05 & 1.39 x 10$^{-15}$ & 1.44 x 10$^{-15}$ \\
        16.10 & 1.42 x 10$^{-15}$ & 1.45 x 10$^{-15}$ \\
        15.26 & 1.45 x 10$^{-15}$ & 1.46 x 10$^{-15}$ \\
        14.49 & 1.48 x 10$^{-15}$ & 1.47 x 10$^{-15}$ \\
        13.80 & 1.51 x 10$^{-15}$ & 1.47 x 10$^{-15}$ \\
        13.18 & 1.53 x 10$^{-15}$ & 1.47 x 10$^{-15}$ \\
        12.61 & 1.56 x 10$^{-15}$ & 1.47 x 10$^{-15}$ \\
        12.08 & 1.59 x 10$^{-15}$ & 1.47 x 10$^{-15}$ \\
        11.60 & 1.62 x 10$^{-15}$ & 1.47 x 10$^{-15}$ \\
        11.15 & 1.65 x 10$^{-15}$ & 1.47 x 10$^{-15}$ \\
        10.74 & 1.67 x 10$^{-15}$ & 1.47 x 10$^{-15}$ \\
        10.36 & 1.70 x 10$^{-15}$ & 1.47 x 10$^{-15}$ \\
        10 & 1.72 x 10$^{-15}$ & 1.46 x 10$^{-15}$ \\
        \hline
        \hline
    \end{tabular}
    \label{tab:fitted_bell_7H}
    \tablefoot{Temperature, in kelvin (K), DFT-computed "Bell" rate constants, in cm$^3$ s$^{-1}$, and "fitted" Bell rate constants, in cm$^3$ s$^{-1}$, obtained by using the fitted parameters ($\alpha$: 6.738 x 10$^{-11}$ cm$^3$ s$^{-1}$, $\beta$: 0.468, $\gamma$: 1733.921 K, $T_{0}$: 198.626 K, RMS: 0.043, R$^{2}$: 0.998) in equation \ref{eq5}.}
\end{table}

\begin{table}
    \centering
    \caption{Bimolecular Eckart rate constants for the 7th hydrogenation of indene.}
    \begin{tabular}{ccc}
        \hline
        \hline
        Temperature & Eckart & Fitted \\
        \hline
        10000 & 3.38 x 10$^{-10}$ & 5.02 x 10$^{-10}$ \\
        282.10 & 5.88 x 10$^{-14}$ & 4.80 x 10$^{-14}$ \\
        143.07 & 1.25 x 10$^{-15}$ & 1.38 x 10$^{-15}$ \\
        95.84 & 1.02 x 10$^{-16}$ & 1.23 x 10$^{-16}$ \\
        72.05 & 2.35 x 10$^{-17}$ & 2.55 x 10$^{-17}$ \\
        57.72 & 9.51 x 10$^{-18}$ & 9.27 x 10$^{-18}$ \\
        48.15 & 5.18 x 10$^{-18}$ & 4.77 x 10$^{-18}$ \\
        41.30 & 3.34 x 10$^{-18}$ & 3.03 x 10$^{-18}$ \\
        36.16 & 2.39 x 10$^{-18}$ & 2.19 x 10$^{-18}$ \\
        32.15 & 1.84 x 10$^{-18}$ & 1.71 x 10$^{-18}$ \\
        28.95 & 1.48 x 10$^{-18}$ & 1.42 x 10$^{-18}$ \\
        26.32 & 1.24 x 10$^{-18}$ & 1.21 x 10$^{-18}$ \\
        24.13 & 1.07 x 10$^{-18}$ & 1.06 x 10$^{-18}$ \\
        22.28 & 9.39 x 10$^{-19}$ & 9.50 x 10$^{-19}$ \\
        20.69 & 8.38 x 10$^{-19}$ & 8.59 x 10$^{-19}$ \\
        19.32 & 7.58 x 10$^{-19}$ & 7.84 x 10$^{-19}$ \\
        18.11 & 6.93 x 10$^{-19}$ & 7.22 x 10$^{-19}$ \\
        17.05 & 6.40 x 10$^{-19}$ & 6.68 x 10$^{-19}$ \\
        16.10 & 5.95 x 10$^{-19}$ & 6.21 x 10$^{-19}$ \\
        15.26 & 5.57 x 10$^{-19}$ & 5.80 x 10$^{-19}$ \\
        14.49 & 5.24 x 10$^{-19}$ & 5.44 x 10$^{-19}$ \\
        13.80 & 4.95 x 10$^{-19}$ & 5.11 x 10$^{-19}$ \\
        13.18 & 4.70 x 10$^{-19}$ & 4.81 x 10$^{-19}$ \\
        12.61 & 4.48 x 10$^{-19}$ & 4.54 x 10$^{-19}$ \\
        12.08 & 4.28 x 10$^{-19}$ & 4.30 x 10$^{-19}$ \\
        11.60 & 4.10 x 10$^{-19}$ & 4.07 x 10$^{-19}$ \\
        11.15 & 3.94 x 10$^{-19}$ & 3.87 x 10$^{-19}$ \\
        10.74 & 3.80 x 10$^{-19}$ & 3.68 x 10$^{-19}$ \\
        10.36 & 3.67 x 10$^{-19}$ & 3.50 x 10$^{-19}$ \\
        10 & 3.55 x 10$^{-19}$ & 3.34 x 10$^{-19}$ \\
        \hline
        \hline
    \end{tabular}
    \label{tab:fitted_eckart_7H}
    \tablefoot{Temperature, in kelvin (K), DFT-computed "Eckart" rate constants, in cm$^3$ s$^{-1}$, and "fitted" Eckart rate constants, in cm$^3$ s$^{-1}$, obtained by using the fitted parameters ($\alpha$: 5.227 x 10$^{-13}$ cm$^3$ s$^{-1}$, $\beta$: 1.974, $\gamma$:  538.782 K, $T_{0}$: 79.126 K, RMS: 0.043, R$^{2}$: 0.999) in equation \ref{eq5}.}
\end{table}

\begin{figure*}
\centering
\includegraphics[width=13cm]{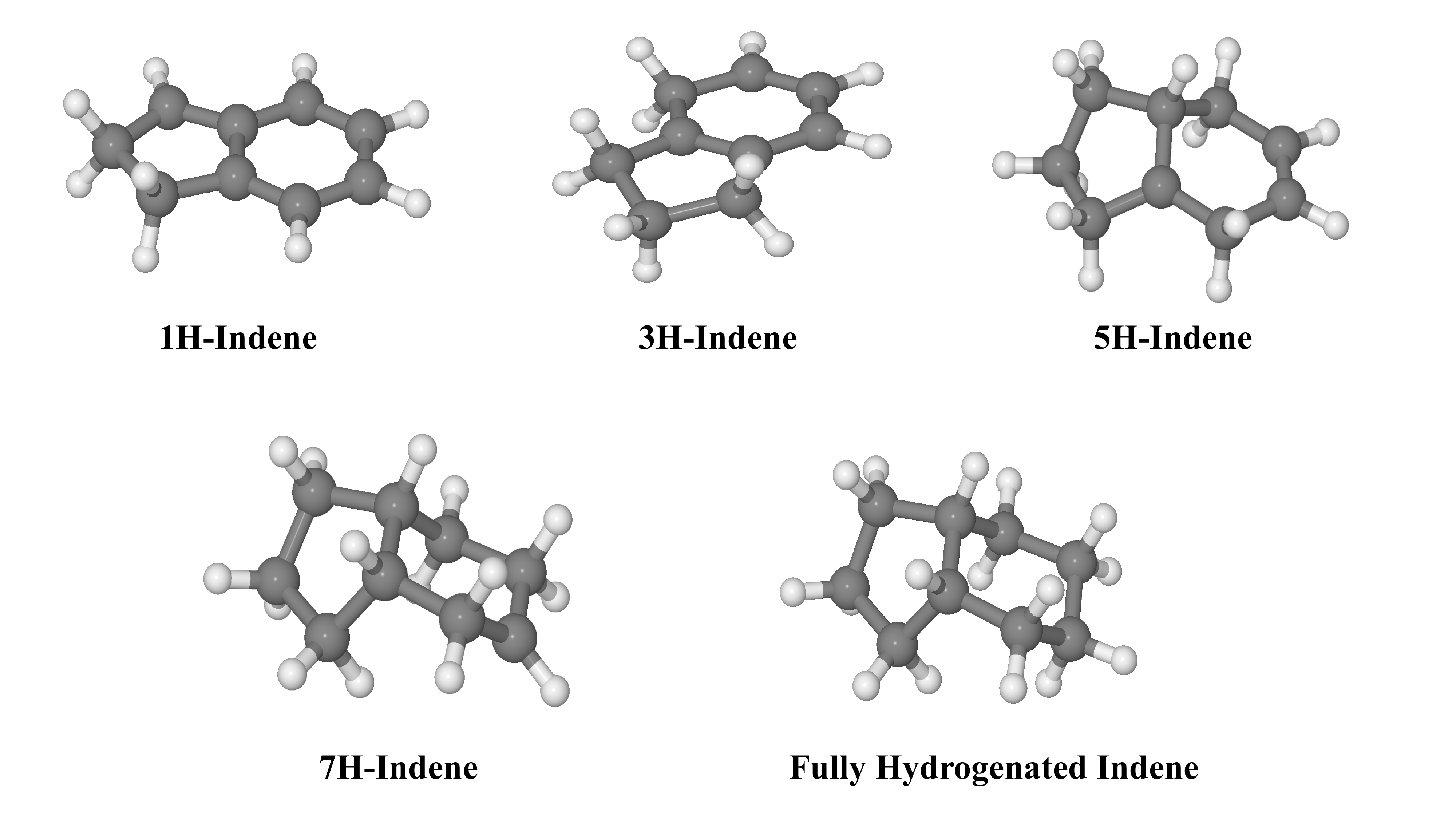}
\caption{Models of odd hydrogenated species of indene using balls and sticks. Gray balls represent carbon atoms, and white balls represent hydrogen atoms.}
\label{Fig:Indene_sequence}
\end{figure*}

\begin{table}
    \centering
    \caption{Bimolecular asymmetric Eckart rate constants for the 1st hydrogenation of indene.}
    \begin{tabular}{lcc}
        \hline
        \hline
        Temperature & Asym. Eckart & Fitted \\
        \hline
        10000 & 5.63 x 10$^{-10}$ & 4.69 x 10$^{-10}$ \\
        282.10 & 8.13 x 10$^{-14}$ & 7.12 x 10$^{-14}$ \\
        143.07 & 1.57 x 10$^{-15}$ & 1.63 x 10$^{-15}$ \\
        95.84 & 1.25 x 10$^{-16}$ & 1.39 x 10$^{-16}$ \\
        72.05 & 3.04 x 10$^{-17}$ & 3.10 x 10$^{-17}$ \\
        57.72 & 1.33 x 10$^{-17}$ & 1.26 x 10$^{-17}$ \\
        48.15 & 7.84 x 10$^{-18}$ & 7.17 x 10$^{-18}$ \\
        41.30 & 5.42 x 10$^{-18}$ & 4.98 x 10$^{-18}$ \\
        36.16 & 4.13 x 10$^{-18}$ & 3.87 x 10$^{-18}$ \\
        32.15 & 3.35 x 10$^{-18}$ & 3.23 x 10$^{-18}$ \\
        28.95 & 2.84 x 10$^{-18}$ & 2.81 x 10$^{-18}$ \\
        26.32 & 2.49 x 10$^{-18}$ & 2.52 x 10$^{-18}$ \\
        24.13 & 2.23 x 10$^{-18}$ & 2.30 x 10$^{-18}$ \\
        22.28 & 2.03 x 10$^{-18}$ & 2.12 x 10$^{-18}$ \\
        20.69 & 1.87 x 10$^{-18}$ & 1.98 x 10$^{-18}$ \\
        19.32 & 1.75 x 10$^{-18}$ & 1.85 x 10$^{-18}$ \\
        18.11 & 1.65 x 10$^{-18}$ & 1.75 x 10$^{-18}$ \\
        17.05 & 1.56 x 10$^{-18}$ & 1.65 x 10$^{-18}$ \\
        16.10 & 1.49 x 10$^{-18}$ & 1.57 x 10$^{-18}$ \\
        15.26 & 1.43 x 10$^{-18}$ & 1.49 x 10$^{-18}$ \\
        14.49 & 1.38 x 10$^{-18}$ & 1.43 x 10$^{-18}$ \\
        13.80 & 1.33 x 10$^{-18}$ & 1.36 x 10$^{-18}$ \\
        13.18 & 1.29 x 10$^{-18}$ & 1.30 x 10$^{-18}$ \\
        12.61 & 1.26 x 10$^{-18}$ & 1.25 x 10$^{-18}$ \\
        12.08 & 1.22 x 10$^{-18}$ & 1.20 x 10$^{-18}$ \\
        11.60 & 1.20 x 10$^{-18}$ & 1.15 x 10$^{-18}$ \\
        11.15 & 1.17 x 10$^{-18}$ & 1.11 x 10$^{-18}$ \\
        10.74 & 1.15 x 10$^{-18}$ & 1.06 x 10$^{-18}$ \\
        10.36 & 1.13 x 10$^{-18}$ & 1.02 x 10$^{-18}$ \\
        10.00 & 1.11 x 10$^{-18}$ & 9.87 x 10$^{-19}$ \\
        \hline
        \hline
    \end{tabular}
     \label{tab:AsymEckart_fit_1H}
     \tablefoot{Temperature, in kelvin (K), DFT-computed “asymmetric Eckart (Asym. Eckart)” rate constants, in cm$^3$ s$^{-1}$, and “fitted” asymmetric Eckart rate constants, in cm$^3$ s$^{-1}$, obtained by using the fitted parameters ($\alpha$: 1.311 x 10$^{-12}$ cm$^3$ s$^{-1}$, $\beta$: 1.695, $\gamma$: 662.929 K, $T_{0}$: 87.512 K, RMS: 0.030, R$^{2}$: 0.999) in equation \ref{eq5}.}
\end{table}

\begin{table}
    \centering
    \caption{Bimolecular asymmetric Eckart rate constants for the 3rd hydrogenation of indene.}
    \begin{tabular}{lcc}
        \hline
        \hline
        Temperature & Asym. Eckart & Fitted \\
        \hline
        10000 & 5.26 x 10$^{-10}$ & 2.03 x 10$^{-10}$ \\
        282.10 & 1.68 x 10$^{-15}$ & 2.91 x 10$^{-15}$ \\
        143.07 & 3.50 x 10$^{-18}$ & 1.24 x 10$^{-17}$ \\
        95.84 & 2.27 x 10$^{-19}$ & 4.69 x 10$^{-19}$ \\
        72.05 & 6.58 x 10$^{-20}$ & 7.80 x 10$^{-20}$ \\
        57.72 & 3.29 x 10$^{-20}$ & 2.93 x 10$^{-20}$ \\
        48.15 & 2.11 x 10$^{-20}$ & 1.69 x 10$^{-20}$ \\
        41.30 & 1.55 x 10$^{-20}$ & 1.22 x 10$^{-20}$ \\
        36.16 & 1.24 x 10$^{-20}$ & 9.93 x 10$^{-21}$ \\
        32.15 & 1.05 x 10$^{-20}$ & 8.66 x 10$^{-21}$ \\
        28.95 & 9.14 x 10$^{-21}$ & 7.85 x 10$^{-21}$ \\
        26.32 & 8.19 x 10$^{-21}$ & 7.29 x 10$^{-21}$ \\
        24.13 & 7.49 x 10$^{-21}$ & 6.87 x 10$^{-21}$ \\
        22.28 & 6.94 x 10$^{-21}$ & 6.52 x 10$^{-21}$ \\
        20.69 & 6.51 x 10$^{-21}$ & 6.23 x 10$^{-21}$ \\
        19.32 & 6.15 x 10$^{-21}$ & 5.98 x 10$^{-21}$ \\
        18.11 & 5.86 x 10$^{-21}$ & 5.75 x 10$^{-21}$ \\
        17.05 & 5.62 x 10$^{-21}$ & 5.54 x 10$^{-21}$ \\
        16.10 & 5.41 x 10$^{-21}$ & 5.34 x 10$^{-21}$ \\
        15.26 & 5.23 x 10$^{-21}$ & 5.15 x 10$^{-21}$ \\
        14.49 & 5.07 x 10$^{-21}$ & 4.97 x 10$^{-21}$ \\
        13.80 & 4.94 x 10$^{-21}$ & 4.81 x 10$^{-21}$ \\
        13.18 & 4.81 x 10$^{-21}$ & 4.65 x 10$^{-21}$ \\
        12.61 & 4.71 x 10$^{-21}$ & 4.49 x 10$^{-21}$ \\
        12.08 & 4.61 x 10$^{-21}$ & 4.35 x 10$^{-21}$ \\
        11.60 & 4.52 x 10$^{-21}$ & 4.21 x 10$^{-21}$ \\
        11.15 & 4.44 x 10$^{-21}$ & 4.07 x 10$^{-21}$ \\
        10.74 & 4.37 x 10$^{-21}$ & 3.94 x 10$^{-21}$ \\
        10.36 & 4.31 x 10$^{-21}$ & 3.82 x 10$^{-21}$ \\
        10.00 & 4.25 x 10$^{-21}$ & 3.70 x 10$^{-21}$ \\
        \hline
        \hline
    \end{tabular}
    \label{tab:AsymEckart_fit_3H}
    \tablefoot{Temperature, in kelvin (K), DFT-computed “asymmetric Eckart (Asym. Eckart)” rate constants, in cm$^3$ s$^{-1}$, and “fitted” asymmetric Eckart rate constants, in cm$^3$ s$^{-1}$, obtained by using the fitted parameters ($\alpha$: 4.255 x 10$^{-13}$ cm$^3$ s$^{-1}$, $\beta$: 1.791, $\gamma$: 1142.702 K, $T_{0}$: 99.832 K, RMS: 0.152, R$^{2}$: 0.995) in equation \ref{eq5}.}
\end{table}

\begin{table}
    \centering
     \caption{{{Bimolecular asymmetric Eckart rate constants for the 5th hydrogenation of indene.}}}
    \begin{tabular}{lcc}
        \hline
        \hline
        Temperature & Asym. Eckart & Fitted \\
        \hline
        10000.0 & 1.41 x 10$^{-10}$ & 9.83 x 10$^{-11}$ \\
        282.10 & 5.60 x 10$^{-14}$ & 8.09 x 10$^{-14}$ \\
        143.07 & 2.57 x 10$^{-15}$ & 3.36 x 10$^{-15}$ \\
        95.84 & 4.35 x 10$^{-16}$ & 4.90 x 10$^{-16}$ \\
        72.05 & 1.71 x 10$^{-16}$ & 1.66 x 10$^{-16}$ \\
        57.72 & 9.97 x 10$^{-17}$ & 8.98 x 10$^{-17}$ \\
        48.15 & 7.05 x 10$^{-17}$ & 6.24 x 10$^{-17}$ \\
        41.30 & 5.53 x 10$^{-17}$ & 4.96 x 10$^{-17}$ \\
        36.16 & 4.62 x 10$^{-17}$ & 4.25 x 10$^{-17}$ \\
        32.15 & 4.02 x 10$^{-17}$ & 3.81 x 10$^{-17}$ \\
        28.95 & 3.59 x 10$^{-17}$ & 3.50 x 10$^{-17}$ \\
        26.32 & 3.28 x 10$^{-17}$ & 3.27 x 10$^{-17}$ \\
        24.13 & 3.04 x 10$^{-17}$ & 3.08 x 10$^{-17}$ \\
        22.28 & 2.85 x 10$^{-17}$ & 2.93 x 10$^{-17}$ \\
        20.69 & 2.70 x 10$^{-17}$ & 2.80 x 10$^{-17}$ \\
        19.32 & 2.57 x 10$^{-17}$ & 2.68 x 10$^{-17}$ \\
        18.11 & 2.46 x 10$^{-17}$ & 2.58 x 10$^{-17}$ \\
        17.05 & 2.37 x 10$^{-17}$ & 2.48 x 10$^{-17}$ \\
        16.10 & 2.30 x 10$^{-17}$ & 2.39 x 10$^{-17}$ \\
        15.26 & 2.23 x 10$^{-17}$ & 2.31 x 10$^{-17}$ \\
        14.49 & 2.17 x 10$^{-17}$ & 2.24 x 10$^{-17}$ \\
        13.80 & 2.12 x 10$^{-17}$ & 2.16 x 10$^{-17}$ \\
        13.18 & 2.07 x 10$^{-17}$ & 2.10 x 10$^{-17}$ \\
        12.61 & 2.03 x 10$^{-17}$ & 2.03 x 10$^{-17}$ \\
        12.08 & 2.00 x 10$^{-17}$ & 1.97 x 10$^{-17}$ \\
        11.60 & 1.96 x 10$^{-17}$ & 1.92 x 10$^{-17}$ \\
        11.15 & 1.93 x 10$^{-17}$ & 1.86 x 10$^{-17}$ \\
        10.74 & 1.90 x 10$^{-17}$ & 1.81 x 10$^{-17}$ \\
        10.36 & 1.88 x 10$^{-17}$ & 1.76 x 10$^{-17}$ \\
        10.00 & 1.86 x 10$^{-17}$ & 1.71 x 10$^{-17}$ \\
        \hline
        \hline
    \end{tabular}
    \label{tab:AsymEckart_fit_5H}
    \tablefoot{Temperature, in kelvin (K), DFT-computed “asymmetric Eckart (Asym. Eckart)” rate constants, in cm$^3$ s$^{-1}$, and “fitted” asymmetric Eckart rate constants, in cm$^3$ s$^{-1}$, obtained by using the fitted parameters ($\alpha$: 1.227 x 10$^{-13}$ cm$^3$ s$^{-1}$, $\beta$: 1.268, $\gamma$: 619.717 K, $T_{0}$: 98.405 K, RMS: 0.052, R$^{2}$: 0.998) in equation \ref{eq5}.}
\end{table}

\begin{table}
    \centering
    \caption{Bimolecular asymmetric Eckart rate constants for the 7th hydrogenation of indene.}
    \begin{tabular}{lcc}
        \hline
        \hline
        Temperature & Asym. Eckart & Fitted \\
        \hline
        10000.0 & 3.37 x 10$^{-10}$ & 3.08 x 10$^{-10}$ \\
        282.10 & 5.47 x 10$^{-14}$ & 4.95 x 10$^{-14}$ \\
        143.07 & 1.20 x 10$^{-15}$ & 1.11 x 10$^{-15}$ \\
        95.84 & 1.23 x 10$^{-16}$ & 1.14 x 10$^{-16}$ \\
        72.05 & 3.63 x 10$^{-17}$ & 3.19 x 10$^{-17}$ \\
        57.72 & 1.80 x 10$^{-17}$ & 1.54 x 10$^{-17}$ \\
        48.15 & 1.14 x 10$^{-17}$ & 9.97 x 10$^{-18}$ \\
        41.30 & 8.34 x 10$^{-18}$ & 7.55 x 10$^{-18}$ \\
        36.16 & 6.60 x 10$^{-18}$ & 6.23 x 10$^{-18}$ \\
        32.15 & 5.52 x 10$^{-18}$ & 5.41 x 10$^{-18}$ \\
        28.95 & 4.79 x 10$^{-18}$ & 4.85 x 10$^{-18}$ \\
        26.32 & 4.27 x 10$^{-18}$ & 4.43 x 10$^{-18}$ \\
        24.13 & 3.88 x 10$^{-18}$ & 4.10 x 10$^{-18}$ \\
        22.28 & 3.58 x 10$^{-18}$ & 3.82 x 10$^{-18}$ \\
        20.69 & 3.34 x 10$^{-18}$ & 3.59 x 10$^{-18}$ \\
        19.32 & 3.15 x 10$^{-18}$ & 3.39 x 10$^{-18}$ \\
        18.11 & 2.99 x 10$^{-18}$ & 3.21 x 10$^{-18}$ \\
        17.05 & 2.86 x 10$^{-18}$ & 3.05 x 10$^{-18}$ \\
        16.10 & 2.75 x 10$^{-18}$ & 2.90 x 10$^{-18}$ \\
        15.26 & 2.65 x 10$^{-18}$ & 2.77 x 10$^{-18}$ \\
        14.49 & 2.57 x 10$^{-18}$ & 2.65 x 10$^{-18}$ \\
        13.80 & 2.49 x 10$^{-18}$ & 2.53 x 10$^{-18}$ \\
        13.18 & 2.43 x 10$^{-18}$ & 2.43 x 10$^{-18}$ \\
        12.61 & 2.37 x 10$^{-18}$ & 2.33 x 10$^{-18}$ \\
        12.08 & 2.32 x 10$^{-18}$ & 2.24 x 10$^{-18}$ \\
        11.60 & 2.27 x 10$^{-18}$ & 2.15 x 10$^{-18}$ \\
        11.15 & 2.23 x 10$^{-18}$ & 2.07 x 10$^{-18}$ \\
        10.74 & 2.19 x 10$^{-18}$ & 1.99 x 10$^{-18}$ \\
        10.36 & 2.16 x 10$^{-18}$ & 1.92 x 10$^{-18}$ \\
        10.00 & 2.13 x 10$^{-18}$ & 1.85 x 10$^{-18}$ \\
        \hline
        \hline
    \end{tabular}
    \label{tab:AsymEckart_fit_7H}
    \tablefoot{Temperature, in kelvin (K), DFT-computed “asymmetric Eckart (Asym. Eckart)” rate constants, in cm$^3$ s$^{-1}$, and “fitted” asymmetric Eckart rate constants, in cm$^3$ s$^{-1}$, obtained by using the fitted parameters ($\alpha$: 1.227 x 10$^{-12}$ cm$^3$ s$^{-1}$, $\beta$: 1.597, $\gamma$: 729.510 K, $T_{0}$: 99.707 K, RMS: 0.034, R$^{2}$: 0.999) in equation \ref{eq5}.}
\end{table}

\begin{table}
    \centering
    \caption{Bimolecular instanton rate constants for the 1st hydrogenation of indene.}
    \begin{tabular}{lcc}
        \hline
        \hline
        Temperature & Instanton & Fitted \\
        \hline
         10000 & 5.63 x 10$^{-10}$ & 5.31 x 10$^{-10}$ \\
        282.10 & 8.26 x 10$^{-14}$ & 6.49 x 10$^{-14}$ \\
        140.00 & 1.26 x 10$^{-15}$ & 1.68 x 10$^{-15}$ \\
        120.00 & 6.52 x 10$^{-16}$ & 7.24 x 10$^{-16}$ \\
        100.00 & 3.98 x 10$^{-16}$ & 2.82 x 10$^{-16}$ \\
        90.00 & 2.58 x 10$^{-16}$ & 1.71 x 10$^{-16}$ \\
        80.00 & 1.04 x 10$^{-16}$ & 1.02 x 10$^{-16}$ \\
        75.00 & 7.78 x 10$^{-17}$ & 7.93 x 10$^{-17}$ \\
        \hline
        \hline
    \end{tabular}
    \label{tab:Instanton_fit_1H}
    \tablefoot{Temperature, in kelvin (K), DFT-computed “Instanton” rate constants, in cm$^3$ s$^{-1}$, and “fitted” instanton rate constants, in cm$^3$ s$^{-1}$, obtained by using the fitted parameters ($\alpha$: 1.020 x 10$^{-12}$ cm$^3$ s$^{-1}$, $\beta$: 1.801, $\gamma$: 620.230 K, $T_{0}$: 99.610 K, RMS: 0.102, R$^{2}$: 0.998) in equation \ref{eq5}.}
\end{table}

\begin{table}
    \centering
    \caption{Bimolecular instanton rate constants for the 3rd hydrogenation of indene.}
    \begin{tabular}{lcc}
        \hline
        \hline
        Temperature & Instanton & Fitted \\
        \hline
        10000. & 5.26 x 10$^{-10}$ & 4.77 x 10$^{-10}$ \\
        282.10 & 1.71 x 10$^{-15}$ & 1.25 x 10$^{-15}$ \\
        195 & 5.01 x 10$^{-17}$ & 7.90 x 10$^{-17}$ \\
        170 & 1.58 x 10$^{-17}$ & 2.52 x 10$^{-17}$ \\
        130 & 3.16 x 10$^{-18}$ & 2.48 x 10$^{-18}$ \\
        100 & 3.16 x 10$^{-19}$ & 2.74 x 10$^{-19}$ \\
        80 & 3.98 x 10$^{-20}$ & 5.29 x 10$^{-20}$ \\
        \hline
        \hline
    \end{tabular}
    \label{tab:Instanton_fit_3H}
    \tablefoot{Temperature, in kelvin (K), DFT-computed “Instanton” rate constants, in cm$^3$ s$^{-1}$, and “fitted” Instanton rate constants, in cm$^3$ s$^{-1}$, obtained by using the fitted parameters ($\alpha$: 8.853 x 10$^{-14}$ cm$^3$ s$^{-1}$, $\beta$: 2.478, $\gamma$: 965.148 K, $T_{0}$: 93.553 K, RMS: 0.136, R$^{2}$: 0.998) in equation \ref{eq5}.}
\end{table}

\begin{table}
    \centering
    \caption{Bimolecular instanton rate constants for the 5th hydrogenation of indene.}
    \begin{tabular}{lcc}
        \hline
        \hline
        Temperature & Instanton & Fitted \\
        \hline
        10000 & 1.41 x 10$^{-10}$ & 1.45 x 10$^{-10}$ \\
        282.10 & 5.67 x 10$^{-14}$ & 3.58 x 10$^{-14}$ \\
        160 & 5.01 x 10$^{-15}$ & 7.60 x 10$^{-15}$ \\
        140 & 3.98 x 10$^{-15}$ & 5.12 x 10$^{-15}$ \\
        100 & 2.00 x 10$^{-15}$ & 1.76 x 10$^{-15}$ \\
        80 & 1.00 x 10$^{-15}$ & 8.03 x 10$^{-16}$ \\
        70 & 5.69 x 10$^{-16}$ & 4.85 x 10$^{-16}$ \\
        50 & 9.58 x 10$^{-17}$ & 1.18 x 10$^{-16}$ \\
        \hline
    \end{tabular}
    \label{tab:Instanton_fit_5H}
    \tablefoot{Temperature, in kelvin (K), DFT-computed “Instanton” rate constants, in cm$^3$ s$^{-1}$, and “fitted” instanton rate constants, in cm$^3$ s$^{-1}$, obtained by using the fitted parameters ($\alpha$: 5.775 x 10$^{-14}$ cm$^3$ s$^{-1}$, $\beta$: 2.235, $\gamma$: 90.910 K, $T_{0}$: 17.432 K, RMS: 0.117, R$^{2}$: 0.996) in equation \ref{eq5}.}
\end{table}

\begin{table}
    \centering
    \caption{Bimolecular instanton rate constants for the 7th hydrogenation of indene.}
    \begin{tabular}{lcc}
        \hline
        \hline
        Temperature & Instanton & Fitted \\
        \hline
        10000 & 3.37 x 10$^{-10}$ & 3.54 x 10$^{-10}$ \\
        282.1 & 5.57 x 10$^{-14}$ & 5.06 x 10$^{-14}$ \\
        150 & 1.58 x 10$^{-15}$ & 1.60 x 10$^{-15}$ \\
        130 & 7.94 x 10$^{-16}$ & 8.69 x 10$^{-16}$ \\
        50 & 1.76 x 10$^{-16}$ & 1.56 x 10$^{-16}$ \\
        \hline
        \hline
    \end{tabular}
    \label{tab:Instanton_fit_7H}
    \tablefoot{Temperature, in kelvin (K), DFT-computed “Instanton” rate constants, in cm$^3$ s$^{-1}$, and “fitted” Instanton rate constants, in cm$^3$ s$^{-1}$, obtained by using the fitted parameters ($\alpha$: 1.195 x 10$^{-11}$ cm$^3$ s$^{-1}$, $\beta$: 1.003, $\gamma$: 1288.304 K, $T_{0}$: 162.831 K, RMS: 0.036, R$^{2}$: 1.000) in equation \ref{eq5}.}
\end{table}

\end{document}